\RequirePackage{fix-cm}
\documentclass[twocolumn]{svjour3}          
\smartqed  
\usepackage{graphicx}
\usepackage{xcolor}
\usepackage{hyperref}

\begin{document}

\title{Synergies of THESEUS with the large facilities of the 2030s and guest observer opportunities
}


\author{
P. Rosati \and
S. Basa \and
A.W. Blain \and
E. Bozzo \and
M. Branchesi \and
L. Christensen \and
A. Ferrara \and
A. Gomboc \and
P.T. O'Brien \and
J.P. Osborne \and
A. Rossi \and
F. Sch\"ussler \and
M. Spurio \and
N. Stergioulas \and
G. Stratta \and
L. Amati \and
S. Casewell \and
R. Ciolfi \and
G. Ghirlanda \and
S. Grimm \and 
D. Guetta \and
J. Harms \and
E. Le Floc'h \and
F. Longo \and
M. Maggiore \and
S. Mereghetti \and
G. Oganesyan \and
R. Salvaterra \and
N.R. Tanvir \and
S. Turriziani \and
S.D. Vergani \and
S. Balman \and
J. Caruana \and
M.H. Erkut \and
G. Guidorzi \and
F. Frontera \and
A. Martin-Carrillo \and
S. Paltani \and
D. Porquet \and
O. Sergijenko
}


\institute{P. Rosati, F. Frontera, C. Guidorzi \at
Department of Physics and Earth Sciences, University of Ferrara, Via G.Saragat, 1 - 44122 Ferrara, Italy \\
\email{rosati@fe.infn.it}   \and
S. Basa \at Aix Marseille Univ, CNRS, CNES, LAM, Marseille, France \and
E. Bozzo \at
Department of Astronomy, University of Geneva, Chemin d'Ecogia 16, CH-1290 Versoix, Switzerland \and
M. Branchesi, J. Harms, S. Grimm, G. Oganesyan \at
 Gran Sasso Science Institute, Viale F. Crispi 7, I-67100 L'Aquila (AQ), Italy; INFN, Laboratori Nazionali del Gran Sasso, I-67100 Assergi, Italy
 \and
A.W. Blain, S. Casewell, P.T. O'Brien, J.P. Osborne, N.R. Tanvir \at
    School of Physics and Astronomy, University of Leicester, University Road, Leicester. LE1 7RH, United Kingdom
 \and             
L. Christensen \at Niels Bohr Institute,
University of Copenhagen, Jagtvej 128, 2200 Copenhagen N, Denmark \and
A. Ferrara \at Scuola Normale Superiore, Piazza dei Cavalieri 7, I-56126 Pisa, Italy \and
A. Gomboc \at Center for Astrophysics and Cosmology, University of Nova Gorica, Vipavska 13, 5000 Nova Gorica, Slovenia \and
L. Amati, G.Stratta, A. Rossi\at
INAF, Osservatorio di Astrofisica e Scienza dello Spazio,  via Piero Gobetti 93/3, 40129 Bologna, Italy \and
F. Sch\"ussler \at IRFU, CEA, Universit\'e Paris-Saclay, F-91191 Gif-sur-Yvette, France \and
M. Spurio \at
Dipartimento di Fisica e Astronomia dell'Universit\`a, Viale Berti Pichat 6/2, 40127 Bologna, Italy;
INFN - Sezione di Bologna, Viale Berti-Pichat 6/2, 40127 Bologna, Italy \and
N.  Stergioulas \at
Department of Physics, Aristotle University of Thessaloniki, 54124 Thessaloniki, Greece \and
R. Salvaterra, S. Mereghetti \at 
    INAF, Istituto di Astrofisica Spaziale e Fisica cosmica,
     via Alfonso Corti 12, 20133 Milano, Italy
\and
R. Ciolfi \at INAF, Osservatorio Astronomico di Padova, Vicolo dell'Osservatorio 5, I-35122 Padova, Italy \and
G. Ghirlanda \at INAF, Osservatorio astronomico di Brera, Via Bianchi 46, I-23807, Merate (LC), Italy       \and
D. Guetta \at ORT Braude, Karmiel, Israel  \and
E. Le Floc'h \at
    AIM, CEA-Irfu/DAp, CNRS, Université Paris-Saclay,
    F-91191 Gif-sur-Yvette, France \and
F. Longo \at Dipartimento di Fisica, Universit\`a degli Studi di Trieste and Istituto Nazionale di Fisica Nucleare, sezione di Trieste, via Valerio 2, I-34127 Trieste, Italy \and
M. Maggiore  \at D\'epartment de Physique Th\'eorique and Center for Astroparticle Physics, Universit\'e de Gen\`eve, 24 quai Ansermet, CH–1211 Gen\`eve 4, Switzerland \and
S. Turriziani \at Physics Department, Gubkin Russian State University, 65 Leninsky Prospekt, Moscow 119991, Russian Federation \and
S. Vergani \at GEPI, Observatoire de Paris, PSL University, CNRS, Place Jules Janssen, 92190 Meudon, France
\and
S. Balman \at
Department of Astronomy and Space Sciences, Istanbul University, Faculty of Science, Beyazit, 34119, Istanbul, Turkey \and
J. Caruana \at Department of Physics and Institute of Space Sciences and Astronomy, University of Malta, Msida MSD 2080, Malta \and
 M.H. Erkut \at
Faculty of Engineering and Natural Sciences, Istanbul Bilgi University
34060, Istanbul, Turkey \and
A. Martin-Carrillo \at
School of Physics and Centre for Space Research, University College Dublin, Dublin 4, Ireland \and
S. Paltani \at Department of Astronomy, University of Geneva, ch. d'Écogia 16, 1290 Versoix, Switzerland
\and
D. Porquet \at Aix Marseille Univ, CNRS, CNES, LAM, Marseille, France \and 
O. Sergijenko \at Astronomical Observatory of Taras Shevchenko National University of Kyiv, Observatorna str., 3, Kyiv, 04053, Ukraine; Main Astronomical Observatory of the National Academy of Sciences of Ukraine, Zabolotnoho str., 27, Kyiv, 03680, Ukraine
}

\date{Received: date / Accepted: date}

\maketitle

\begin{abstract}
The proposed THESEUS mission will vastly expand the capabilities to monitor the high-energy sky. It will specifically exploit large samples of gamma-ray bursts to probe the early universe back to the first generation of stars, and to advance multi-messenger astrophysics by detecting and localizing the counterparts of gravitational waves and cosmic neutrino sources. The combination and coordination of these activities with multi-wavelength, multi-messenger facilities expected to be operating in the 2030s will open new avenues of exploration in many areas of astrophysics, cosmology and fundamental physics, thus adding considerable strength to the overall scientific impact of THESEUS and these facilities. We discuss here a number of these powerful synergies and guest observer opportunities. 
\keywords{multi-messenger astrophysics \and gamma-ray bursts \and X-ray sources \and gravitation wave sources \and neutrino sources}
\end{abstract}

\section{Introduction}
\label{intro}

Major discoveries in astrophysics and cosmology over the last decades have benefited greatly from the synergy among space and ground-based facilities over the entire electromagnetic spectrum. Wide area surveys have played a key role in this endeavor by probing large population of astrophysical sources and often discovering the rarest objects in the Universe, feeding follow-up studies with large narrow-field facilities from the radio to the X-ray bands. Some of these facilities have themselves conducted deep surveys pushing the exploration to the early universe, albeit over small fields (for example the HST and Chandra deep fields), thus stimulating multi-wavelength, multi-observatory observations in the same sky regions. 

Recent history has also shown that some of the most interesting and often unexpected breakthroughs come from the discovery of transient phenomena, when monitoring capabilities are combined with large area surveys. This has been particularly true in the high-energy (HE) domain, where gamma-ray bursts (GRBs), the most luminous known sources of electromagnetic radiation, have emerged over the last decade as powerful probes of the distant universe, by pinpointing the collapse of massive stars (the long duration GRBs), potentially back to the very first stars, a few hundred million years after the Big Bang. The other class of short duration GRBs has now been securely associated with neutron star (NS) compact binary mergers, after the detection of gravitational waves (GW) by the Ligo-Virgo interferometers (GW170817, \cite{LVC-BNS}), and the prompt identification of the electro-magnetic counterpart from $\gamma$-rays to radio wavelengths has ushered in the new era of multi-messenger astrophysics.

In general, long-term monitoring of the high-energy sky reveals time variability of galactic and extra-galactic sources of various kind. Together with the most energetic transient phenomena, one has unique access to extreme physical processes, with a strong bearing on fundamental physics: strong gravity regime, nuclear physics at extreme densities, most efficient energy production mechanisms and particle acceleration, extreme magnetic fields, the cosmic synthesis of heavy metals.

\begin{figure*}
\centering 
 \includegraphics[width=0.8\textwidth]{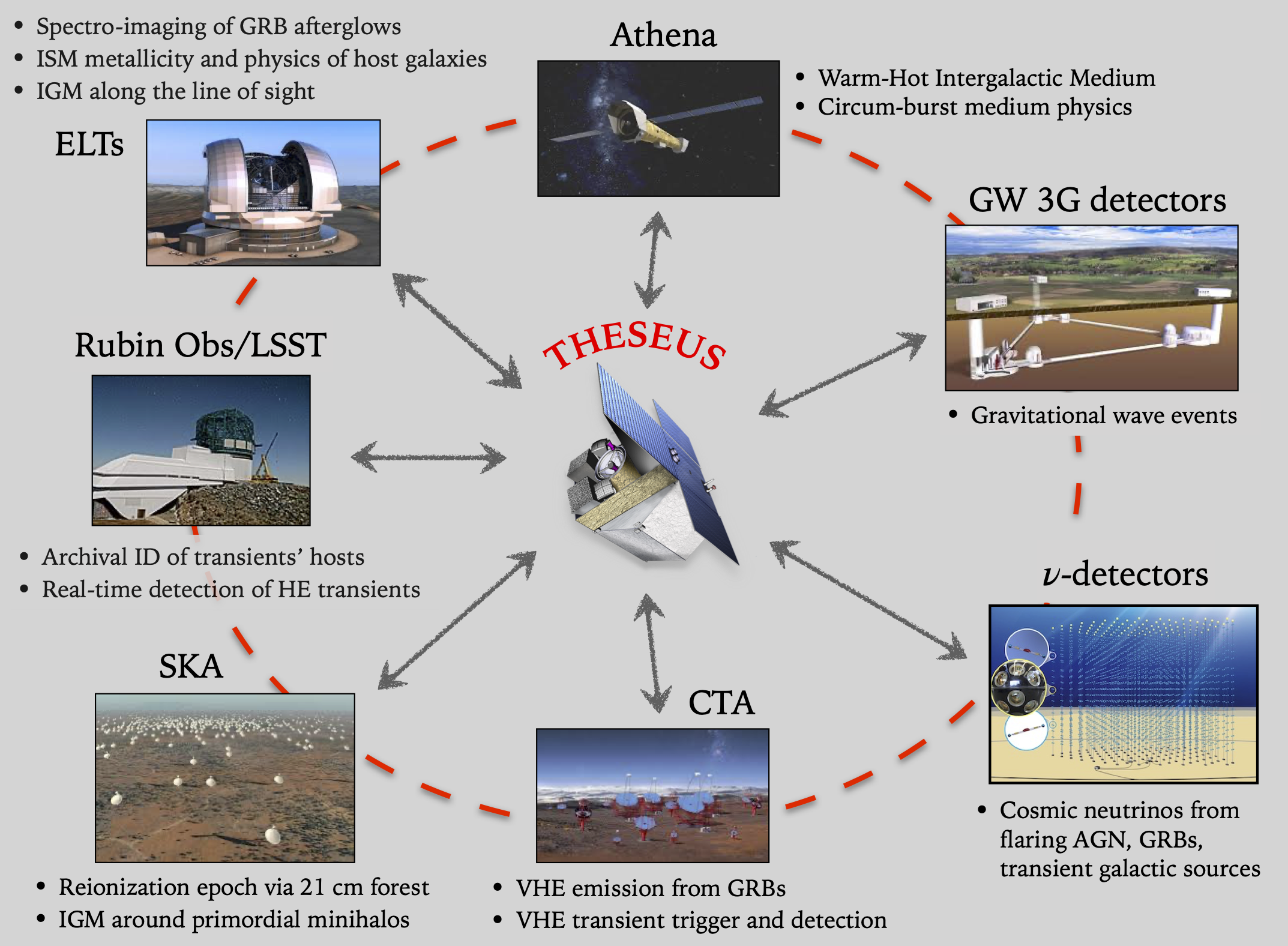}
\caption{THESEUS will work in synergy on a number of themes (bullets) with major multi-messenger facilities in the 2030s and will provide targets and triggers for follow-up observations with several of these facilities.}
\label{fig:Overview}      
\end{figure*}

The proposed mission THESEUS (Transient High-Energy Sky and Early Universe Surveyor)\footnote{We refer to the THESEUS Assessment Study Report (\url{https://sci.esa.int/s/8Zb0RB8}) for an overview of the mission, a description of the on-board instrumentation (XGIS, SXI, IRT), and the key scientific objectives. The mission profile and science cases are also discussed in detail in accompanying papers \cite{Amati2021,Mereghetti2021,Ciolfi2021,Tanvir2021,Ghirlanda2021}.}
is ideally posed to play a pivotal role in the 2030s in the exploration of transient high-energy sky, at the time when major multi-wavelength, multi-messenger facilities will be operating. These include extremely large telescopes (ELTs) and Rubin/LSST in the optical-NIR, the X-ray Athena observatory, the next generation GW interferometers and neutrino detectors, the Square Kilometre Array (SKA), TeV telescopes (e.g. CTA) (Fig.~\ref{fig:Overview}).

Thanks to the wide and deep monitoring of the high-energy sky (SXI+XGIS instruments from 0.3 keV to 10 MeV), with focusing capabilities in the soft X-ray (SXI), and an on-board 70 cm infrared telescope (IRT), THESEUS will feed these powerful facilities with GRBs and other selected transient events, for dedicated follow-up observations; THESEUS however will also respond to triggers from these facilities. The combination of such a vast network of complementary information 
will unleash the full scientific potential of the mission and expand the scientific impact of all these future facilities. 
We discuss below specific science synergies between THESEUS and such major facilities.

\section{The Athena X-ray mission}
\label{sec:Athena}

Athena (Advanced Telescope for High Energy Astrophysics) is an ESA-led X-ray space mission, planned to be launched in the early 2030s, which has a number of primary science requirements exploiting high-energy transients to probe physical questions. These include: probing stars in the early universe; using GRBs as backlight to characterize the warm-hot intergalactic medium (WHIM); and probing galactic and extra-galactic variable sources, such as tidal disruption events (TDE), Active Galactic Nuclei (AGN) and stellar binary systems. For events with a sufficiently small positional uncertainty, Athena can also contribute significantly to the area of multi-messenger astrophysics. To maximise the role of Athena, it is therefore essential that it promptly receives information on the location, brightness, redshift and object class for a newly discovered transient. Athena must in some cases be provided with targets very quickly (within a few hours) while others can be provided on timescales of days – weeks. THESEUS is ideally posed to deliver this information to Athena.

Extragalactic transients are sufficiently rare that intrinsically large field-of-view instruments are needed to identify sufficiently sizeable samples of bright objects. The Swift-BAT and the Fermi-GBM lack on-board redshift determination capability, and critically, they are unlikely to be still operational in the Athena era as they were launched in 2004 and 2008 respectively. Although the upcoming transient discovery missions SVOM and Einstein Probe could still be operational (albeit beyond their design lifetimes), they lack multi-wavelength redshift determination capability on-board. Only the multi-wavelength THESEUS, due for launch very close in time to Athena, can find transients, determine redshift on-board, conduct a soft X-ray survey and find multi-messenger counterparts in the X-ray and IR, while rapidly communicating discoveries to other facilities. 
Despite being independent missions, the combination of Athena and THESEUS in the early 2030s would greatly enhance the science return of both. Two prominent Athena science objectives ideally match the capabilities of THESEUS, by requiring very rapid (target sent to Athena within a few hours) identification of bright GRBs with redshift determination: 

\begin{figure}
  \includegraphics[width=\linewidth]{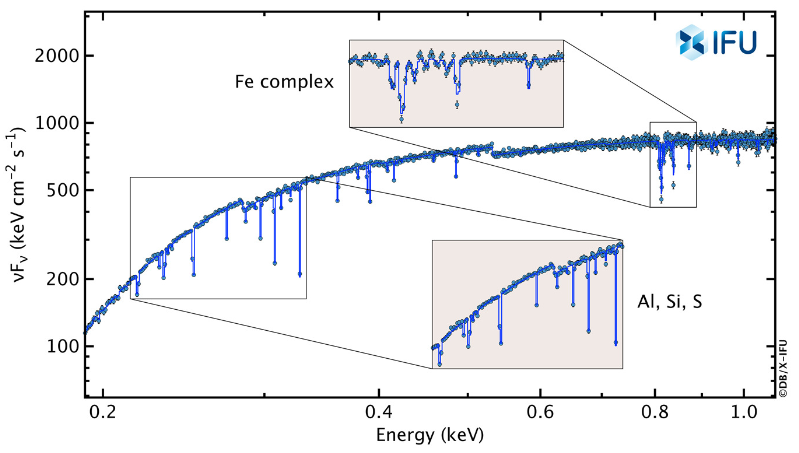}
\caption{A simulated Athena/X-IFU spectrum of a medium bright (fluence: $4\times 10^{-7} {\rm erg~cm^{-2}}$) $z=7$ GRB afterglow characterized by deep resonant lines from the ISM of the GRB host galaxy. An effective intrinsic column density of $2\times 10^{22} {\rm cm^{-2}}$ was assumed [credit: Athena/X-IFU Consortium].}
\label{fig:Athena}  
\end{figure}

\begin{enumerate}
\item
Probe the first generation of stars (Cosmic Dawn), the formation of the first black holes, the dissemination of the first metals, and the primordial IMF. This is to be achieved by determining the elemental abundances of the medium around high-redshift GRBs. The Athena requirement is to observe 25, $z>7$ GRBs (Fig.~\ref{fig:Athena}). 

\item
Measure the local cosmological baryon density in the WHIM to better than 10\% and constrain structure formation models in the local density regime by measuring the redshift distribution and physical parameters of WHIM filaments. The Athena requirement is to observe 100 filaments towards bright GRBs up to $z=1$.
\end{enumerate}

To enable a statistically reliable study of the period of cosmic dawn with Athena requires raising the current high-redshift GRB detection rate by more than an order of magnitude, a requirement well met by THESEUS but well beyond the capability of any other current or approved future mission. Athena and THESEUS also have similar fields of regard, so prompt follow-up is achievable. 
If followed-up within half a day, a 50~ksec exposure using the Athena/X-IFU will typically contain hundreds of thousands of photons from the brightest high-redshift GRBs.  Under the requirement that the GRB afterglow observed by Athena/X-IFU should have over a million photons in a 50\,ksec observation, exploiting THESEUS GRB capabilities, a recent study found that the number of expected WHIM aborbers during the 4-year nominal lifetime of the Athena would range between 45\,--\,137 \cite{Walsh2020}. This shows that the rate of GRB discovery from THESEUS would readily provide the required rate of very bright low-redshift GRBs for the Athena WHIM science requirements. 

Athena has multiple science requirements that assume the availability of astrophysical high-energy transients which it can follow-up on timescales of days to weeks depending on the object class. Examples include: (a) studying the nature of stellar disruption and subsequent accretion onto super-massive black holes during TDEs, (b) observing stellar-mass and super-massive black holes in both quiescence and outburst phases to probe the accretion process, (c) probing disk-corona systems in changing-look AGN and (d) studying the counterparts to multi-messenger (GW and neutrino) events. 

To this end, it is critical to monitor large areas of sky in real-time to know the current accretion state of either known or new examples of such systems. These Athena science requirements are highly synergistic with those of THESEUS, which has a primary science objective to perform an unprecedented real-time, high-cadence, deep monitoring of the X-ray transient universe in order to identify a wide variety of extragalactic and galactic transients and to identify the counterparts to multi-messenger sources. The two monitors on THESEUS will simultaneously observe the sky over an energy range exceeding a factor of five orders of magnitude and will be particularly sensitive to the softest energies down to 0.3~keV, providing a perfect complement to the Athena observing bandpass. The THESEUS/IRT can also provide complementary data for source monitoring and classification. THESEUS will detect tens of TDEs per year, many magnetars/soft-gamma Repeater (SGRs), supernovae shock breakouts, supergiant fast X-ray transient (SFXTs), thermonuclear bursts from accreting neutron stars, Novae, dwarf novae, stellar flares, AGN and blazars \cite{Mereghetti2021}. The data will be available within days from the THESEUS data processing system. THESEUS will fly in the era when the next generation of gravitational-wave and neutrino detectors will provide routine detections. Counterparts of such events found by THESEUS will have location accuracies far superior to those of the multi-messenger observatories and will be observable by Athena in a single pointing.

\section{Extremely Large Telescopes}
\label{sec:ELT}

The THESEUS mission will operate in the decade when the next generation of giant, 20 to 40m class telescopes will be fully operational, specifically, the Giant Magellan Telescope (GMT, 22-m equivalent diameter), the Thirty Meter Telescope (TMT) and the 39-m ESO Extremely Large Telescope (E-ELT). Dedicated follow-up observations of selected GRB events discovered by THESEUS with these facilities will realise and enhance the full scientific goals of the mission, particularly in the exploration of the early universe and in the study of the optical/near-infrared (NIR) counterparts of GW signals coming from NS-NS and NS-BH mergers associated with short GRBs (SGRB). Over the last two decades, we had a preview of the powerful synergy between GRB-dedicated experiments, such as {\it Swift}/BAT, {\it Fermi}/GBM, and dedicated follow-up campaigns with 8-10m class telescopes, for example with the characterization of the interstellar medium in a few GRB hosting galaxies in the first billion years of cosmic history ($z>6$) with VLT spectroscopy (e.g. \cite{Hartoog2015}), or with observations of the kilonova associated with the GW170817 binary NS merger (e.g. \cite{Pian2017}). The synergy between THESEUS and the giant optical/NIR telescopes of the 2030s, assisted with laser guide Adaptive Optics (AO), is expected to be far more powerful, thus enabling transformative science. For this reason, following up these GRBs are high-priority research goals for many of the planned ELT instruments.  
Deep follow up observations with the largest ground-based facilities will exploit the drastic improvement of their spectro-imaging sensitivity and angular resolution compared to current 10m-class telescopes. They will also take advantage of the two order of magnitude increase in the number of GRBs at any redshift, and particularly one order of magnitude more at $z>6$ (likely even beyond redshift 10). In addition, THESEUS on-board NIR spectroscopy capabilities will ensure arcsec accurate location and redshift estimates of the GRBs, thus allowing time-critical follow-up observations of pre-selected targets to be scheduled with the appropriate instrumentation on these facilities. 

We refer to Tanvir et al. 2021 (\cite{Tanvir2021}, this volume) for a detailed description of a range of critical scientific cases related to the high-redshift universe, including simulations and forecasts on expected performances when E-ELT follow-up observations of THESEUS GRBs at $z>6$ are carried out. Briefly, long duration GRBs at high redshift are generated by the collapse of massive UV-bright stars and can therefore trace star formation in the early universe, in a complementary (and likely unbiased) fashion when compared to the conventional UV rest-frame selection of Lyman-break or Ly-$\alpha$ emitting galaxies in deep HST surveys. Using GRBs as signposts of their host galaxies, these can be unveiled even in the very low-luminosity regime (e.g. \cite{McGuire2016}) and their redshift measured with ELTs (if only a photometric redshift with a 10\% uncertainty can be measured with the IRT onboard THESEUS). Moreover, the bright power-law emission of the GRB afterglow illuminates the interstellar and intergalactic medium, which can be investigated in absorption with high-dispersion, high-S/N spectroscopy afforded by these future facilities. This spectral analysis provides detailed chemical abundance patterns and dust content of the GRB host galaxy, even in the low-metallicity regime, and can reveal evidence of Pop III chemical enrichment. Afterglow spectroscopy also yields an accurate measurement of the neutral hydrogen column density of the host galaxy that can be disentangled from a component of a (partly) neutral intergalactic medium (IGM), and finally an estimate of the average escape fraction of ionizing radiation. A knowledge of the physical properties of primordial galaxies well into the reionization epoch, together with the measured GRB rate, will provide an independent determination of the early rise of the star formation rate in the Universe, indicating whether UV emitting sources at the very faint end of the luminosity function can drive the reionization at $z>6$. 

\begin{figure}
  \includegraphics[width=\linewidth]{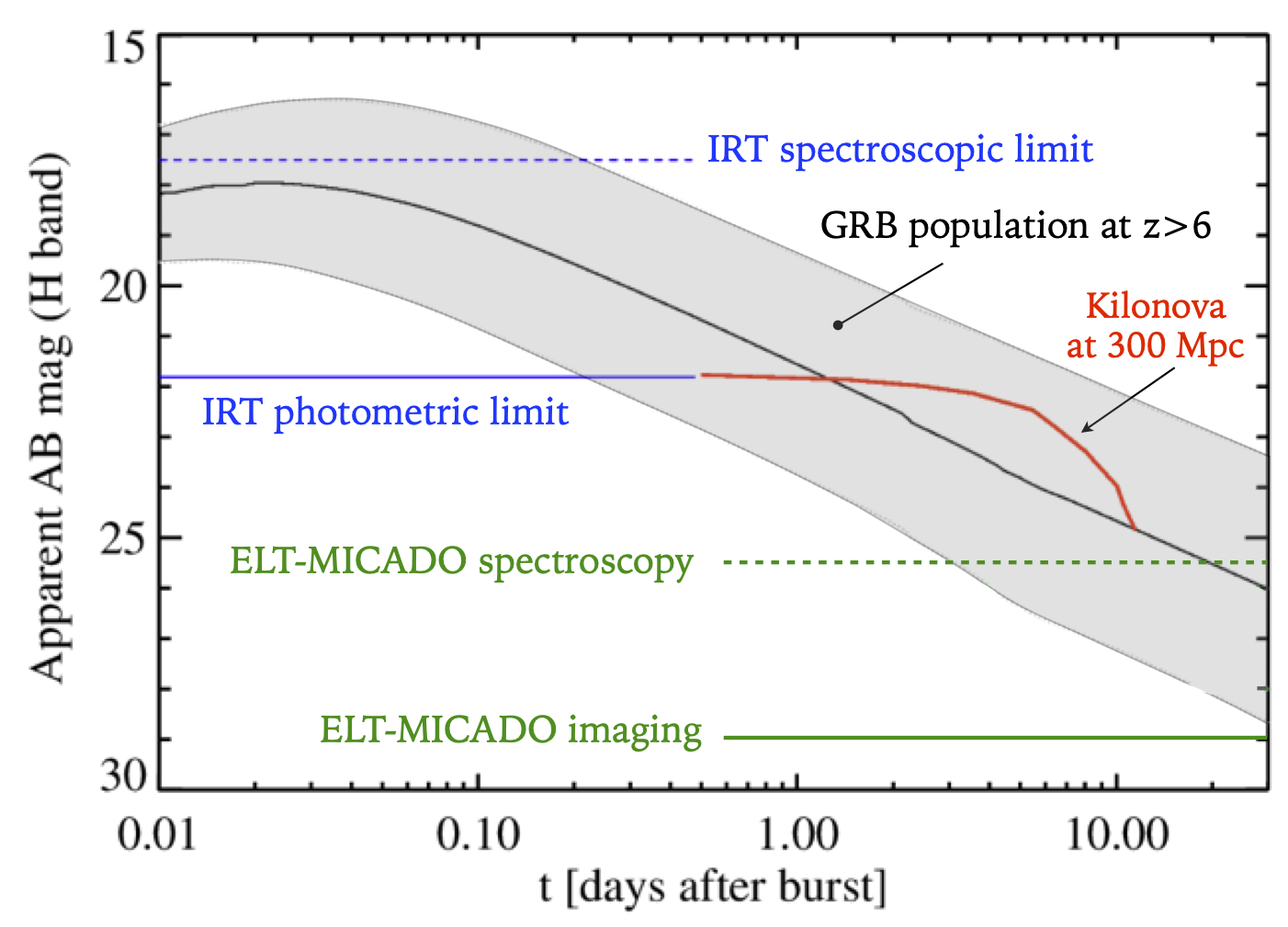}
\caption{Expected H-band light curves of the afterglow of high redshift GRBs and a kilonova from NS-NS merger at 300 Mpc. The shaded area indicates the median (black line) and 1-$\sigma$ scatter of 100 simulated light curves drawn from GRB population models at $z > 6$ following  \cite{Ghirlanda2015}. The red curve represents the kilonova associated to GW170817, projected to a distance of 300 Mpc, using H-band photometry from \cite{Rossi2020}. Horizontal lines indicate the limiting magnitude of THESEUS IRT (600s and 1800s exposure for photometric and spectroscopic limit, with SNR$=5$ and 3, respectively,) and ESO ELT-MICADO (1 h exposure with SNR$=5$) in imaging and spectroscopic mode respectively
(re-adapted from \cite{Maiorano2018}).}
\label{fig:kilonova}  
\end{figure}

We refer to Ciolfi et al. 2021 (\cite{Tanvir2021}, this volume), and section~\ref{sec:GW} below, for a discussion of the transforming capabilities of THESEUS in detecting and localizing  the electromagnetic counterparts of gravitational waves. Also in this case, follow-up observations with the E-ELT, in particular with the first light instrument MICADO+MAORY \cite{MAORY}, can lead to a characterization of the NS-NS and NS-BH binary mergers, and the host galaxies of the associated SGRB event, at distances significant larger than GW170817. In Fig.~\ref{fig:kilonova}, we show representative NIR lightcurves of the afterglow of high-z GRBs and a kilonova from a binary NS merger associated with a SGRB at 300 Mpc. The imaging and spectroscopic capabilities of the E-ELT enable a characterization of such events over a prolonged period after the burst, significantly extending the observations with the IRT on-board THESEUS. 

An interesting new avenue opened by high spatial resolution observations with MICADO-MAORY, with extreme AO yielding diffraction limited performance at $\sim 6$ mas resolution, is the possibility to characterize the stellar population of the site of the SGRB explosion from which the progenitor originated. Similarly, the $50\,\mu as$ astrometric capabilities of MICADO can be used, in principle, to study the jet and expanding envelope of the SGRB afterglow in nearby events.

\section{The Square Kilometre Array (SKA) and other ground-based radio facilities}
\label{sec:SKA}

The Square Kilometer Array (SKA) will be the largest radio observatory, expected to be fully operating in the 2030s, consisting of two arrays of thousands of dishes, one operating at mid-to-high frequencies located in South Africa, one at low radio frequencies located in Australia (e.g. \cite{SKA2013}).
The SKA will drive transformational science in many fields of astronomy.   
In particular, it will enable an ideal technique to study the evolution of cosmic reionization via the measurement of the 21~cm radiation from neutral hydrogen atoms (due to the hyperfine structure of the triplet and the singlet levels of the hydrogen ground state). The 21~cm sky contains fluctuations around the mean (“global”) signal, which encode information on the physical state of hydrogen, largely representative of all baryons, in the Dark Ages and in the Epoch of Reionization (EoR) \cite{Mitra_Reion}.

\begin{figure}
  \includegraphics[width=\linewidth]{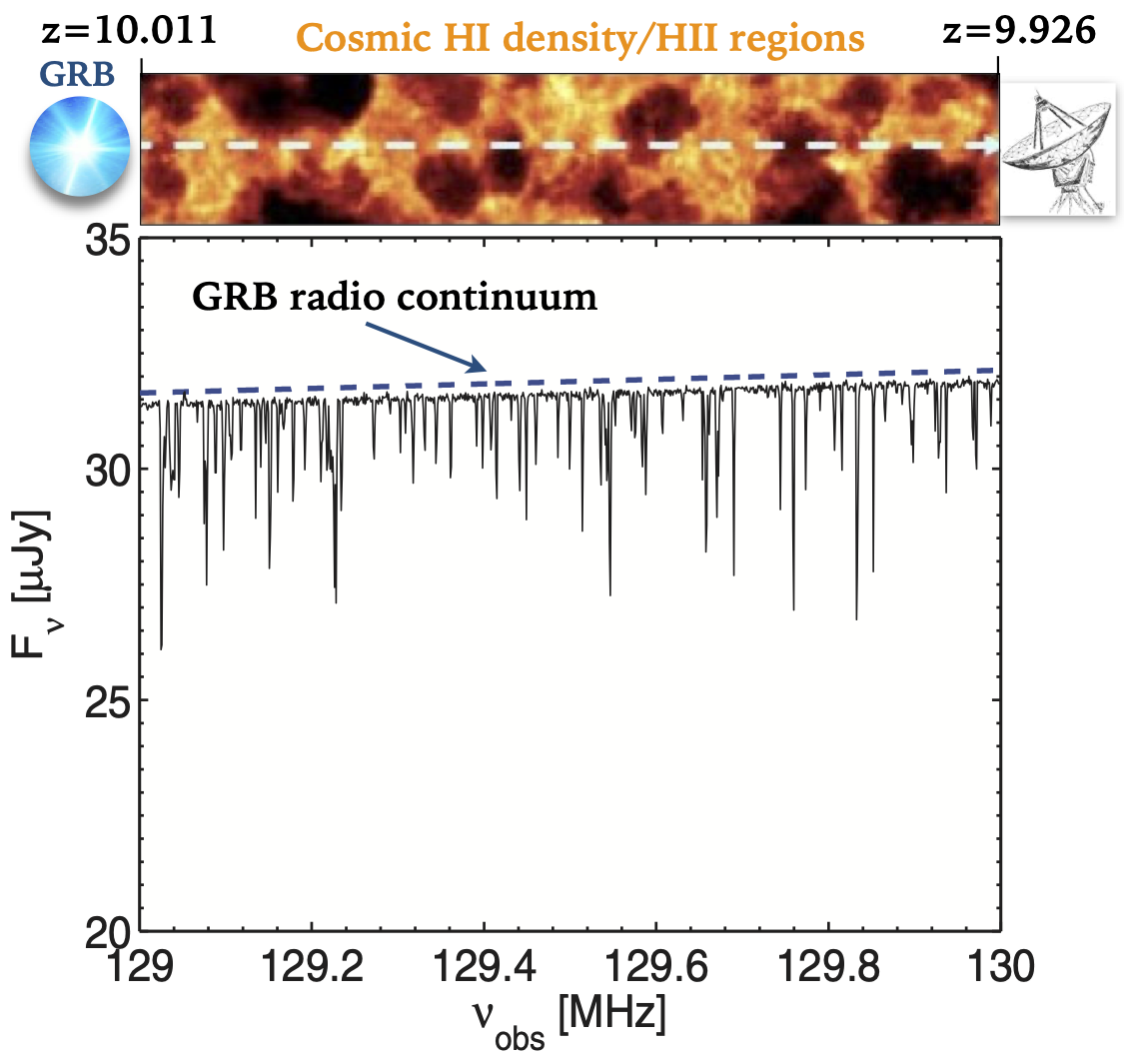}
\caption{Representation of the 21~cm absorption lines (known as the ``21~cm forest'') produced by non-linear structures during the early stage of reionization \cite{Xu2011}.}
\label{fig:SKA}  
\end{figure}

Canonical tools to study reionization include the polarization of the Cosmic Microwave Background, and most relevant here, absorption line spectra of bright sources, typically quasars, located in the reionization epoch. However, the paucity of these objects at early epochs, particularly in the first billion years of cosmic history, has prevented a comprehensive study of the EoR. The large sample of high-z long GRBs from THESEUS will open a complementary way to explore the reionization process avoiding the proximity effects. This has been made possible for the first time after the detection of GRBs at $z >5$ \cite{Gehrels2004-Swift}, and specifically demonstrated by the studies by \cite{Totani06} and \cite{Gallerani08}, who have used these sources to constrain the ionization state of the IGM at high redshift by modelling its optical afterglow spectrum. 

An even more exciting prospect is provided by 21~cm radio observations of GRB afterglows, particularly with SKA. This idea was first explored by \cite{Xu2011} who investigated the 21~cm absorption lines (known as the “21~cm forest'') produced by non-linear structures during the early stage of reionization, i.e. the starless minihaloes and the dwarf galaxies (see Fig.~\ref{fig:SKA}). The infalling gas velocity around minihaloes/dwarf galaxies strongly affects the line shape and, with the low spin temperatures outside the virial radii of the systems, gives rise to horn-like line profiles. The authors compute synthetic spectra of 21~cm forest for the radio afterglows of GRBs. Broadband observation against GRB afterglows can also be used to reveal the evolving 21~cm signal from both minihaloes and dwarf galaxies. The number, strength and clustering of 21~cm absorption lines depend very sensitively on the intensity of the X-ray background produced by e.g. high-mass X-ray binaries or early black holes. This experiment will then offer a unique opportunity to explore cosmic dawn and the rise of the first structures, including black holes. The synergy between THESEUS and SKA will be fundamental to achieve these ambitious goals.

The SKA in its final configuration is expected to detect several GRB radio afterglows \cite{Ghirlanda2013} within a follow-up program based on external triggers like those provided by THESEUS for both long and short GRBs. These observations are expected to provide fundamental insight on the physics of jets powering GRBs. Radio observations at early times can reveal the reverse shock emission component,which bears information on the magnetization of the outflow (e.g. GRB 130427 \cite{Perley2014}, GRB161219B \cite{Laskar18}).

Radio imaging with micro-arcseconds astrometric precision will allow SKA to measure the source proper motion out to $z\sim 0.3$, depending on source intrinsic parameters and flux density, thus constraining the viewing angle (e.g. \cite{Fernandez2021}) through multi--wavelength modelling of the afterglow emission (e.g. \cite{Mooley2018,Ghirlanda2019}). The high resolution and sensitivity of the SKA will also help to unveil off-axis jets and orphan GRBs (\cite{2015A&A...578A..71G,Marcote2019}).

For $\sim 50$\% of the bursts, SKA can perform radio calorimetry  \cite{Ghirlanda2013}, after the trans-relativistic transition occurring $\sim 100$ days after the explosion when the afterglow emission should be 0.1--10 $\mu$Jy, providing an estimate of the true, collimation corrected energetics of the GRBs \cite{Shivvers2011}. 

Finally, radio observations of high-$z$ GRBs detected by THESEUS could also provide a route to distinguish GRBs powered by Pop III stars. The large expected GRB energetics (e.g. $10^{54-57}$ erg) from these progenitors, if powered by magneto-rotational energy of the black hole and/or intense disk neutrino flux, and the possible relatively small external medium density (e.g. $n<0.1$ cm$^{-3}$) would produce a quite bright radio afterglow with peak flux densities in the range 100$\mu$Jy at 100 days to 10 mJy at $10^3$ days.  Indeed, such events are expected to produce brighter radio afterglows peaking at much later times than standard Pop II/I bursts \cite{Burlon2016}.

\subsection{The Next Generation Very Large Array (ngVLA)}
\label{ngvla}

The Next Generation Very Large Array (ngVLA) is a project under development at the National Radio Astronomy Observatory (NRAO). This new radio facility builds upon the scientific and technical legacy of both the Jansky VLA and ALMA with the aim to dramatically improve our understanding of planets, galaxies, black hole and the dynamic radio sky \cite{Murphy+2018}.

The ngVLA is being designed to observe in the 1.2-116 GHz range, with up to 20 GHz of instantaneous sampled bandwidth. The main array will consist of 214 antennas, each 18m in diameter, complemented by a short baseline array of 19 antennas, each 6m in diameter, and a long baseline array of 30 antennas, each 18m in diameter. The antennas in the main array will provide baselines ranging from tens of meters up to 1000 km, thus achieving mas-resolution, whereas the long baseline array will extend the maximum baseline to $\sim$~9000 km, enabling sub-mas imaging capabilities. The short baseline array instead will increase the sensitivity at the larger angular scales which cannot be detected with the main array.
With this design, the ngVLA will reach 10 times the sensitivity of the Jansky VLA and ALMA, and its dense core at km-baselines will provide high surface brightness sensitivity \cite{Selina+2018}. Therefore, the ngVLA will bridge ALMA and SKA capabilities, opening a new window in studies of thermal lines, continuum emission and polarimetric imaging. 

The high-resolution and fast-mapping capabilities of ngVLA will allow the identification of the (eventual) radio counterparts of the high energy transient sources discovered by THESEUS. In such a case, THESEUS alerts can be used as an external trigger for ngVLA.

 For example, the ngVLA will be an ideal instrument to study the radio emission components of short GRBs: the frequency range would allow for distinguishing better between the different components of short GRBs emission; in particular, a rapid follow-up at radio ($<$ 1 day) with ngVLA would give the possibility to detect the reverse shock emission \cite{LR2018}. Other components which can be probed with the ngVLA are early-time self-absorbed forward shock emission, later-time optically thin forward shock and off-axis jet emission, and combined with additional multi-wavelength observation would allow us to constrain the relevant physical parameters.

Moreover, the ngVLA can perform multi-frequency radio follow-up observations of tidal disruption events (TDEs) \cite{vanVelzen+2018} in the 3-100 GHz range, with the possibility to detect thermal flares up to z$\sim$0.2. Furthermore, coordinated X-ray/radio observations of higher luminosity non-thermal TDEs at higher redshift could also be performed, especially in order to establish the true nature of sources detected in blind surveys.  

Mereghetti et al. 2021 (\cite{Mereghetti2021}, this volume) describe how THESEUS can detect new X-ray-transient AGN in the local universe and blazar flares, monitor them, and trigger a series of multi-wavelength follow-up observations. Follow-ups with the ngVLA will permit to image exceedingly faint radio emission on scales of 10s to 100s of milliarcseconds, and allow for studying the formation, structure, and evolution of jets \cite{Lister+2018}. Cross-correlation studies of emission at different wavelengths are crucial to probe, e.g. the connection of multiple emitting regions in blazars' jets.

\section{The Rubin Observatory Legacy Survey of Space and Time}
\label{sec:LSST}

The Vera Rubin Observatory will carry out the Legacy Survey of Space and Time (LSST), a 10-year long survey which will create a multi-color, dynamic view of the sky \cite{LSST_book}. With its large field of view of 9.6 deg$^2$, it will cover around 10000 deg$^2$ each night, in six optical bands (u, g, r, i, z and y), thus mapping the entire visible Southern sky in just a few nights. With its 8.4 m diameter primary mirror, it will reach very faint magnitudes (up to 24.4 in r band) in a single exposure.
The survey will produce a catalog of around 37 billion objects: 20 billion galaxies, 17 billion stars and orbits for 6 million bodies in the Solar System.

Science operations of the LSST project are planned to start in 2024 and continue through the next decade. Its four main science drivers are: (1) understanding dark matter and dark energy, (2) hazardous asteroids and the remote Solar System (3) formation and structure of the Milky Way, and (4) variable and transient sources. 

In particular, the LSST will be a revolutionary and powerful transient machine: a stream of 1-10 million of time-domain events per night (among them tens of thousands of new transients) are expected to be detected with real-time data analysis. They will be publicly distributed through Alert Brokers within 60 seconds of observation. Currently, a number of LSST observing strategies are still under study, with different combinations of cadence and sky coverage, to provide adequate discovery space for a variety of classes of transients. In general, the LSST will therefore be complementary to THESEUS’s survey capabilities for high-energy transient phenomena. 

The catalog of 20 billion galaxies provided by the LSST (combined magnitudes r $< 27.5$ and time resolved measurements to r $< 24.5$) – characterized in shape, color, and variability – will be an invaluable resource for identification studies of THESEUS transients, including location and properties of the host galaxies of GRBs, GW events, TDEs etc. and their galactic environments. 
It has been estimated that the LSST will observe in total tens of millions of variable stars \cite{2014ApJ...796...53R}, with other estimates up to $\sim$ 135 million \cite{LSSTAlerts}. 
Estimates show that the discovery rate of cataclysmic variables could be as high as several 1000 per night \cite{2014ApJ...796...53R}, while the discovery rate of active galactic nuclei (AGNs) will be $\sim$ 3000 per night in the beginning of the LSST survey, and will decrease by a factor of 50 over first 4 years of the survey. In total the LSST will produce a catalog of millions of AGNs \cite{LSSTAlerts}.

The LSST catalogs of transients will clearly be of particular interest for THESEUS. It is very likely that LSST observations, with emphasis on variable and transient sources, will continue after the first 10 years of operations and that LSST will continue to generate (as a conservative estimate) thousands of transient alerts of interest to high-energy scientists every night. The success rate of such observations will depend on the adopted LSST cadence strategy.

Contemporaneous observations by the LSST in the optical and by THESEUS in IR and X-rays, will, for example, be very beneficial for studies of TDEs (see also \cite{Mereghetti2021}). 
According to estimates by \cite{2020ApJ...890...73B} the LSST will detect around 3500-8000 new TDEs per year, providing solid constraints on the rate of these events.
Combined X-ray and optical observations of TDE events will ultimately improve our understanding on accretion processes in black holes.  

Although LSST observing strategies under consideration are not designed for GRB afterglow detection, an appreciable number of GRBs afterglows is detectable - of the order of about 25 events per year \cite{LSSTObsStrategy}. The LSST could also detect 50 orphan afterglows per year, more than any other planned survey \cite{2015A&A...578A..71G}.

Triggering and follow-up observations with the LSST are currently not envisaged, however unique data (detections or upper limits) will be available in the LSST images taken (serendipitously) just prior, during or soon after transient events detected by THESEUS. Therefore, the LSST database can be searched for information on optical counterparts and host galaxies corresponding to transients detected by THESEUS. 
Furthermore, once the major science goals of LSST will be reached in about ten years, observational time dedicated to target of opportunity programs for highly interesting targets, such as gravitational-wave signals, are highly likely.

Synergy flow is possible also in the other direction: interesting transients discovered by the LSST could be observed/followed up with THESEUS Guest Observer (GO) program (see Sec.\ref{sec:GOObs}). This could be applied also in observations of Solar System objects. For example, the LSST will make a large number of comet observations: about 10$^4$ comets will be observed on average of 50 times by the LSST during its main survey \cite{2010PhDT.......241S}. In addition to ground-based observations of particularly interesting cases, a range of cometary emission and absorption features could be observed with THESEUS IR spectroscopy.

\section{Next Generation Gravitational Wave detectors}
\label{sec:GW}

\subsection{Ground-based GW observatories}

\begin{figure}
\centering 
 \includegraphics[width=\linewidth]{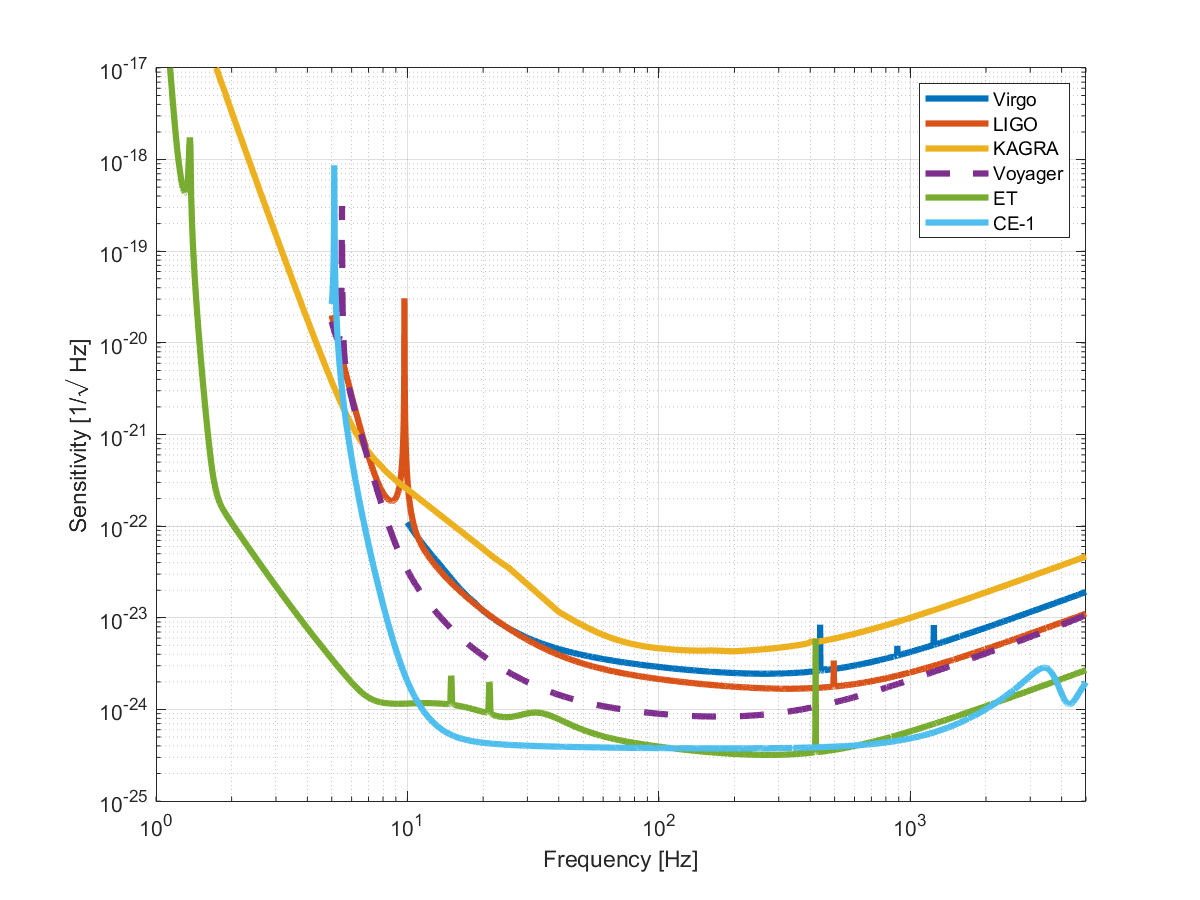}
\caption{Strain sensitivities as a function of frequency for the next generation of GW observatories, ET \cite{ETsensitivity} and CE Phase I \cite{Reitze:2019iox} with respect to the sensitivities of the upgraded instruments of the existing LIGO, Virgo and Kagra expected to be operational in 2025 \cite{LVK-LRR-2020-short}, and LIGO Voyager \cite{Adhikari2020}, a possible major upgrade of the LIGO inteferometers in the late 2020s.}
\label{fig:ET-CE}      
\end{figure}

The next generation ground-based gravitational wave observatories in the 2030s, such as Einstein Telescope (ET) \cite{Punturo:2010zz,2020JCAP...03..050M} and Cosmic Explorer (CE) \cite{Evans:2016mbw,Reitze:2019iox}, will come online with sensitivities about an order of magnitude better than today’s detectors, and with an observation band extended to lower frequencies \cite{GWIC2020}. They will be an unprecedented resource to observe compact object mergers across cosmic history back to the early universe. They will offer the opportunity to explore the origin, evolution, and demography of binary neutron-star (BNS) and neutron star-black hole (NS-BH) mergers. Planned upgrades of the current gravitational-wave detectors will bridge the second generation to the third generation interferometers \cite{LVK-LRR-2020-short,GWIC2020}. For LIGO, a major upgrade in the late 2020s, LIGO Voyager, would provide a significant increase in sensitivity, and at the same time it will implement some key technologies needed as well for the ET and CE \cite{Adhikari2020}. Voyager is expected to increase the detection range by a factor of about 3 over Advanced LIGO, thus bringing the event rate from several hundreds to thousands of detections.
Fig.~\ref{fig:ET-CE} shows the expected sensitivities of ET and CE, together with the sensitivities of LIGO, Virgo and Kagra which include technological upgrades expected to be deployed in 2025 \cite{LVK-LRR-2020-short}, as well as LIGO Voyager. 

\begin{figure}
\centering 
\includegraphics[width=\linewidth]{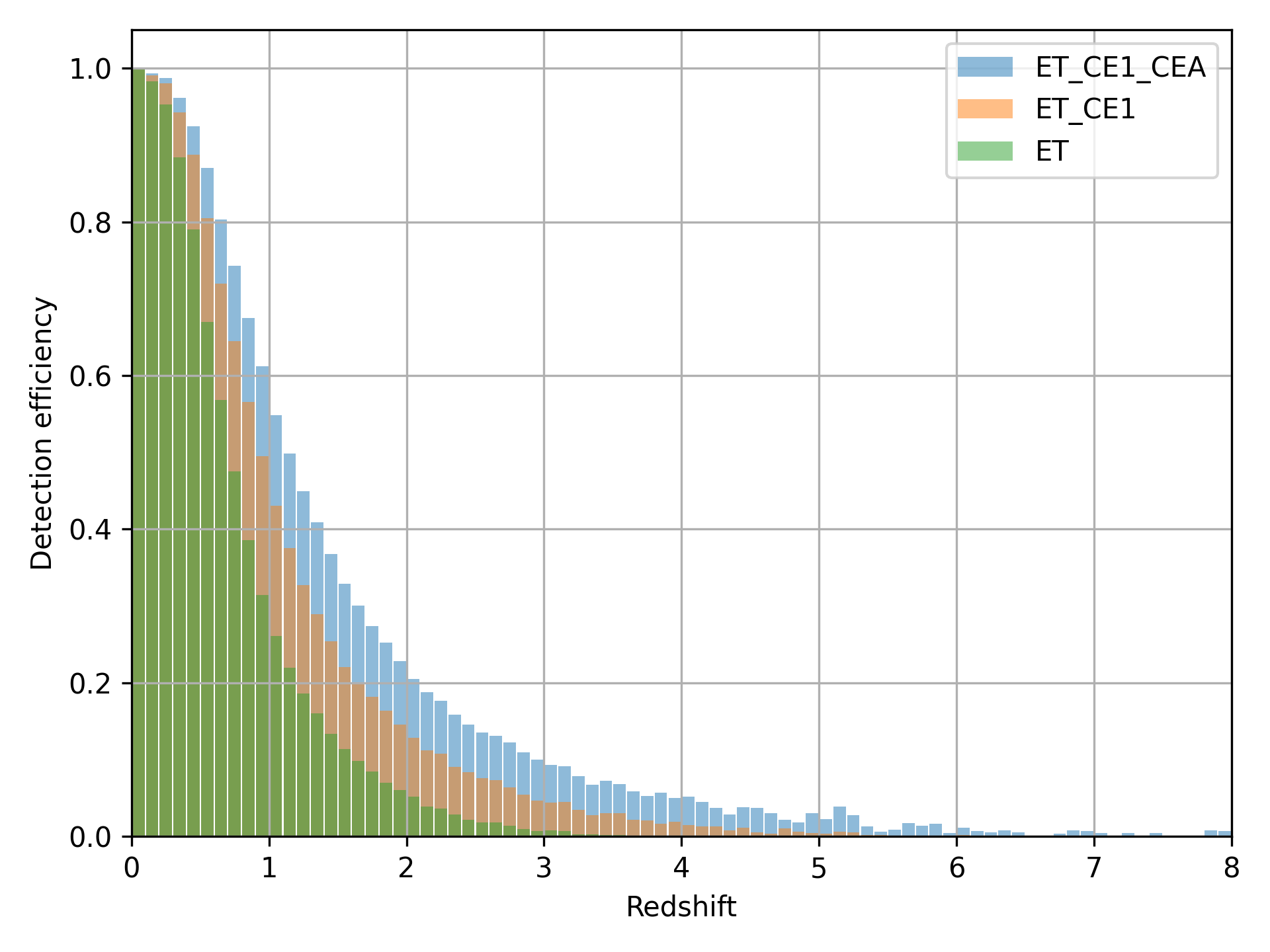}
\caption{Detection efficiency for binary neutron star mergers by ET alone, ET operating with CE located in USA, and ET operating with one CE in USA and one CE in Australia. Detection SNR threshold set equal to 8.}
\label{fig:ETCEdeteff}      
\end{figure}

Fig.~\ref{fig:ETCEdeteff} shows the detection efficiency for an astrophysical population of binary neutron stars randomly oriented and located in space with component masses fixed at $1.4\, M_{\odot}$, with a detection threshold corresponding to a signal-to-noise ratio (SNR) of 8. ET will be able to detect  binary neutron-star mergers up to a redshift $\sim 3$, with a 50\% detection efficiency at $z=0.8$.  We also consider two networks of third generation (3G) detectors, one composed of ET and CE (Phase I) in the USA (ET+CE), and one also including a second CE in Australia. The horizon redshift extends to about 5 (8), with 50\% detection efficiency corresponding to $z= 1\, (1.2)$ for ET+CE (ET+2CE).

Operating in synergy with the upgraded second generation detectors as well as ET and CE, THESEUS will probe the electromagnetic emission from the multi-component outflows in connection with gamma-ray bursts. The wide field of view of THESEUS will enable the prompt detection of the X-ray emission associated with the BNS and NSBH mergers. A longer time monitoring of the X-ray emission will unveil the properties of the relativistic jets as well as the role of the merger remnant on the properties of the afterglow emission, i.e., the extended emission and the X-ray plateaus. In \cite{Ciolfi2021},  we  have evaluated the joint detection rate to be of few tens per year for short  gamma-ray  bursts in synergy with the third generation (considering both jet aligned and misaligned from the observers). This estimate can increase to orders of 100 joint detections per year when considering the short GRB extended emission, the off-axis high latitude emission from the jet \cite{Ascenzi2020}, and the soft X-ray emission from the spin-down of new-born neutron stars. 

\begin{figure}
\centering 
\includegraphics[width=\linewidth]{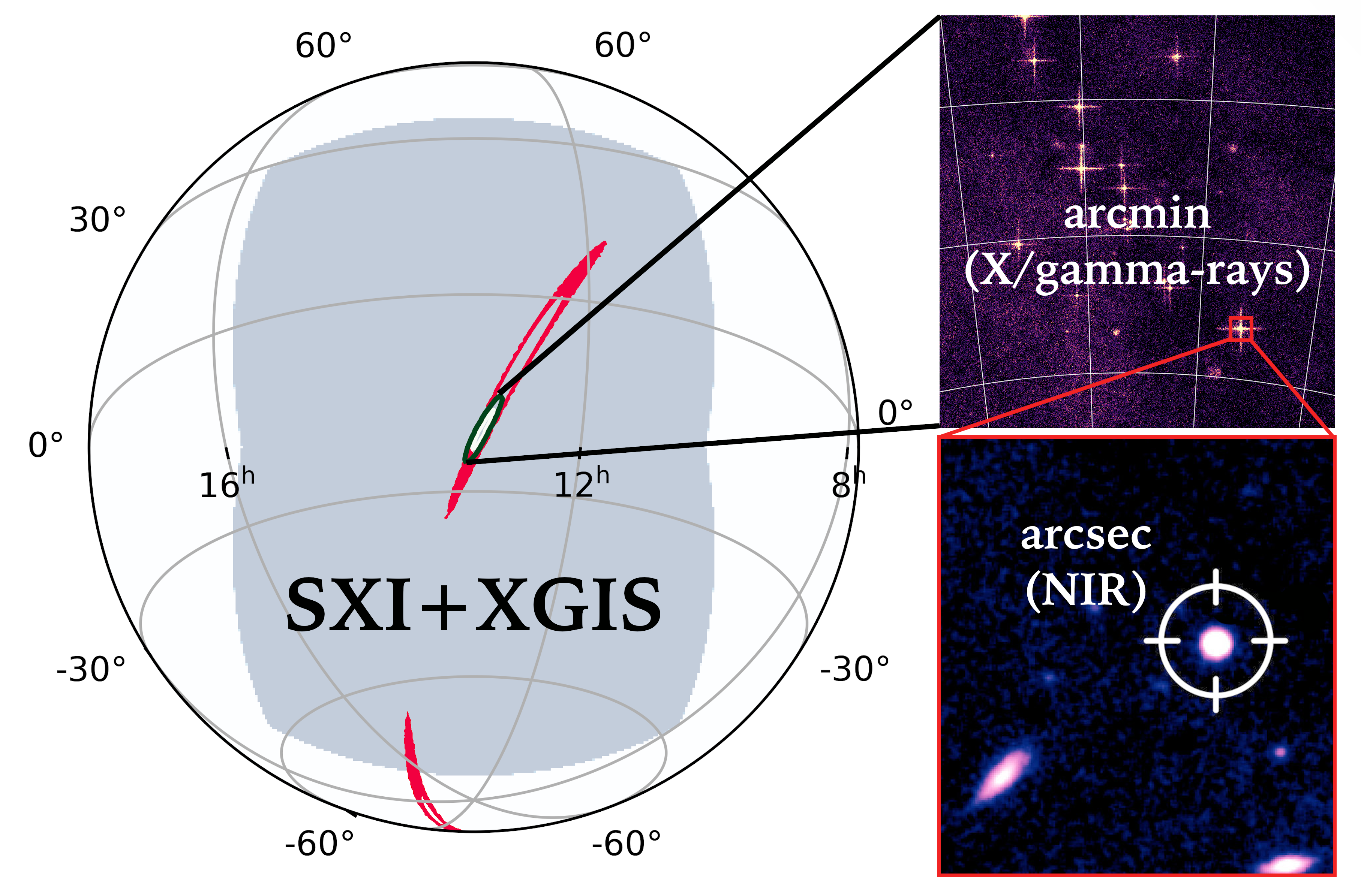}
\caption{THESEUS can cover with a single exposure most of the sky uncertainty regions that will be obtained  from GW source detection with both 2G and 3G interferometers (red contours), allowing independent detection of the electromagnetic counterpart and accurate sky localization down to arcmin/arcsec level, exploiting the SXI and IRT instruments.}
\label{fig:Theseus_loc}      
\end{figure}

THESEUS will perform an unprecedented deep monitoring of the transient universe, and represents the major asset during the 2030s to localize and identify the electromagnetic counterparts to GW detections from BNS and NS-BH mergers at high-redshift. Among the $10^5$ BNS mergers expected to be detected each year (based on \cite{2021MNRAS.502.4877S} and assuming a NS mass of 1.4 M$_{\odot}$), ET alone will able to localize only a fraction of about 1\% of them, with an error box smaller than 100 deg$^2$ up to a redshift of 0.3\footnote{Localization uncertainties obtained from the same population used to estimate the detection efficiency of Fig.~\ref{fig:ETCEdeteff} and considering a detection SNR $>8$.}. 
An operating CE-ET network would localize 50\% of BNS mergers within $10 \, {\rm deg}^2$ at a distance of only $z=0.5$, while it would still detect sources which are at significantly larger luminosity distances \cite{PhysRevD.97.104064}. For the events with a jet pointing to us or slightly misaligned, THESEUS will represent a unique instrument to detect and localize the associated short GRB and enable prompt ground-based follow-up observations, as illustrated in Fig.~\ref{fig:Theseus_loc}.  

While the gravitational wave observables bring information about the merger components, i.e., their masses and spins, as well as the properties of the merger remnant, the electromagnetic observables diagnose the characteristics of the outflows launched from the merger. Therefore, joint detections provide the unique possibility to link the properties of the progenitor to the large energy released in the electromagnetic channel, and are a powerful tool to answer fundamental questions about the nature of short GRBs: what is the jet launching mechanism and its efficiency? Are there  any  systematic  differences  between  NS-BH  and NS-NS jets? Do NS-NS/NS-BH jets have universal structure? What is the nature of merger remnant and what is its connection with afterglow features? 

BNS mergers that do not lead to a prompt collapse to a black hole, produce remnants in which fluid modes are excited. The post-merger GW spectrum is dominated by a quadrupolar fluid oscillation at a frequency $f_{\mathrm{peak}}$ (in the range of a few kHz) \cite{PhysRevLett.94.201101}, with additional nonlinear contributions \cite{Stergioulas2011,2015PhRvD..91l4056B}. The sensitivity of the LIGO and Virgo GW detectors was not sufficient to detect the post-merger phase of the BNS merger GW170817  \cite{LVC-BNS,2017ApJ...851L..16A}, or the likely BNS merger GW190425 \cite{2020ApJ...892L...3A}. 
However, post-merger GW detections are expected to be achieved by 3G detectors \cite{2014PhRvD..90f2004C,2019PhRvD.100j4029B,2019PhRvD..99d4014T,2020PhRvD.102d3011E,2020PhRvL.125z1101H}. 

A confirmed absence of post-merger GW emission would imply a prompt collapse to a BH. This is expected to be accompanied by systematically less massive and more neutron-rich ejecta, resulting in a less luminous and redder kilonova than for the case of a delayed collapse \cite{Margalit2019,2020IJMPD..2941015F}. The accurate sky coordinates that THESEUS will promptly disseminate, will allow for kilonova follow-up campaigns with large facilities such as the ELTs and the verification of these predictions. 
The threshold on the total mass of the BNS system which determines the prompt collapse to a black hole depends mainly on the equation of state (EOS) and the component mass ratio. Empirical relations connecting these properties have been found through numerical simulations \cite{2020PhRvL.125n1103B,2020arXiv201004461B}. With a sufficient number of post-merger GW observations with 3G detectors, the fraction of BNS mergers that result in a prompt collapse vs. those that result in a delayed (or no) collapse will be determined. This information would have direct implications for the fraction of GRBs which have peculiar features on the prompt and afterglow emission, as for instance the short GRB "extended emission" or the X-ray afterglow "plateaus" (see more in \cite{Ciolfi2021}). Joint detections of confirmed prompt- or delayed collapse BNS mergers  with  3G  detectors  and  THESEUS  would  significantly advance our understanding of the emission mechanisms powering short GRBs. 

The joint 3G/THESEUS detections will also have a significant impact on cosmology. It will be possible to build a statistical  significant sample of  standard  sirens  to  measure the Hubble constant with a precision around 1\% \cite{Ciolfi2021}.  Furthermore, ET and CE will reach distant astrophysical sources and make precise measurements of the expansion rate, sensitive to the dark energy density and modifications of General Relativity (GR) on cosmological scales (see \cite{Maggiore2020},
which describes the science cases of ET). Modified gravity theories predict deviations from GR in the propagation of GWs, which make luminosity distances measured from the GW amplitude different from the standard (electromagnetic) luminosity distances. While this possible distance difference complicates the search for an electromagnetic counterpart when the GW distance is used as prior for the search, the temporal coincidence (maintained in the theories where the speed of gravity is equal to speed of light as proved for GW170817) makes short GRBs the ideal counterpart to be detected and used to test deviations from GR. For almost all modified gravity models, the deviations from GR can be parametrized in terms of two parameters $(\Xi_0,n)$  as~\cite{Belgacem:2018lbp}
\begin{equation}
\frac{d_L^{\rm gw}(z)}{d_L^{\rm em}(z)}=\Xi_0+\frac{1-\Xi_0}{(1+z)^n},
\end{equation}
where $d_L^{\rm em}$ is the electromagnetic luminosity distance.

\begin{figure}\centering
\includegraphics[width=0.48\textwidth]{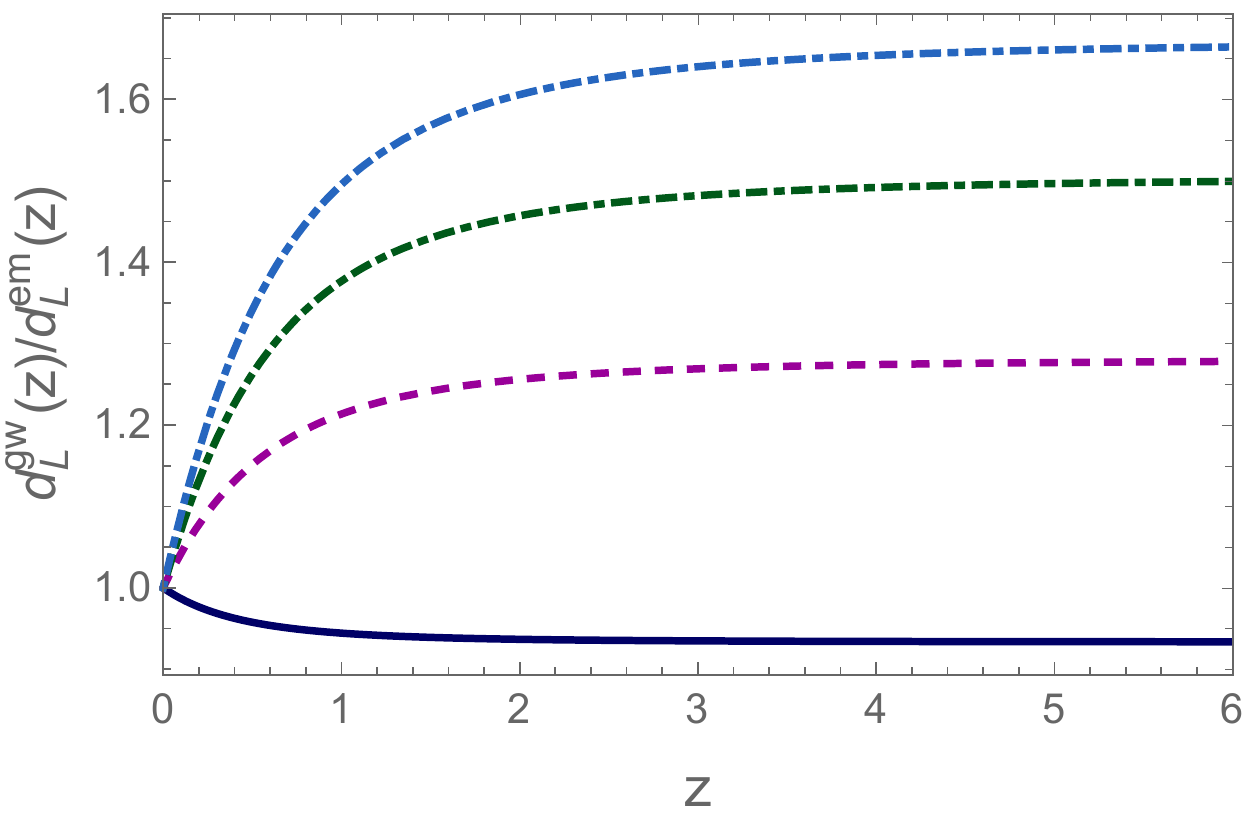}
\caption{Example of the ratio of the gravitational to electromagnetic luminosity distance in a modified gravity model, for different values of a parameter related to the  energy scale of inflation. In the upper curve,  the deviations of $d_L^{\rm gw}(z)$ from $d_L^{\rm em}(z)$ reaches $60\%$ at large redshift (from ~\cite{Belgacem:2019lwx}). \label{fig:xi0w0}
}
\end{figure}

With a few hundreds of joint electromagnetic-GW detections, ET will constrain $\Xi_0$ to below 1\% or better \cite{Belgacem:2019lwx}, but large deviations might be detected even with just a single THESEUS-ET detection. An example of deviations from GR, measured by ratio between the GW and electromagnetic luminosity distance is shown in Fig.~\ref{fig:xi0w0} for a specific model of modified gravity.

ET will access the full population of stellar-mass binary black-hole (BBH) and intermediate mass BH mergers across the cosmic history of stellar evolution (see e.g. \cite{Maggiore2020}). Thus, ET will probe the earliest populations of the stars in the Universe, and at the same time will test the existence of primordial black-holes predicted to form in the early universe. ET will be able to detect BBH mergers at redshift larger than 30, where BBH from astrophysical origin are not expected (see e.g. \cite{PBH2020}). THESEUS is sensitive to very distant long GRBs, possibly probing Pop III stars. ET and THESEUS can therefore work as multi-probe instruments to unveil the first structures of the Universe, and disentangle astrophysical Pop III stars from primordial BHs.

\subsection{Space-based GW detectors}
The Laser Interferometer Space Antenna (LISA) is an ESA-led mission expected to be launched in 2034, which will extend the gravitational wave spectrum to the low frequency range, from about 10$^{-4}$ Hz up to 1  Hz \cite{LISA}. The LISA mission is composed of three spacecrafts that will form a triangular configuration with an arm length of 2.5 million km. It will open a new window by detecting gravitational waves from more massive and/or wider orbit objects, compared to the ground-based gravitational detectors, such as supermassive black hole (SMBH) mergers (with masses in the range 10$^4$-10$^6$\,M$_{\odot}$), extreme mass ratio inspirals, and ultra-compact binaries in our Galaxy. 
If THESEUS and LISA will operate in the same years, the FoV of THESEUS will be very valuable to detect promptly the counterpart of gravitational events that could have several to 10s degrees error box positions. This will also trigger target of opportunity observations with other ground-based telescopes and space missions in the 2030s.  
Even in the case that there will be no observational time overlap between the two missions, THESEUS could also play an important role for LISA. In fact, it might be able to detect promising astrophysical sources for LISA, namely, by detecting the modulated X-ray emission expected for SMBH binaries which are very close to merger (from year to day time-scales). If their X-ray flux is high enough to be detected by the SXI instrument, THESEUS could therefore provide, in advance, positions of SMBH binary candidates to be scrutinized by LISA and with multi-wavelength observational campaigns. 

Furthermore, THESEUS and LISA will provide  complementary visions on the early epoch in the Universe when the first collapsed structures emerged, the former through identification of the farthest galaxies pinpointed by GRBs, the latter through the merging of SMBH possibly out to $z\sim20$. 

\section{Neutrino detectors}
\label{sec:neutrinos}

Understanding the physical origin of cosmic high-energy diffuse neutrinos, discovered by the IceCube experiment \cite{Aarsten2013}, has become one of the most pressing question of modern astrophysics and fundamental physics. 

\begin{figure*}
\centering 
\includegraphics[width=0.8\textwidth]{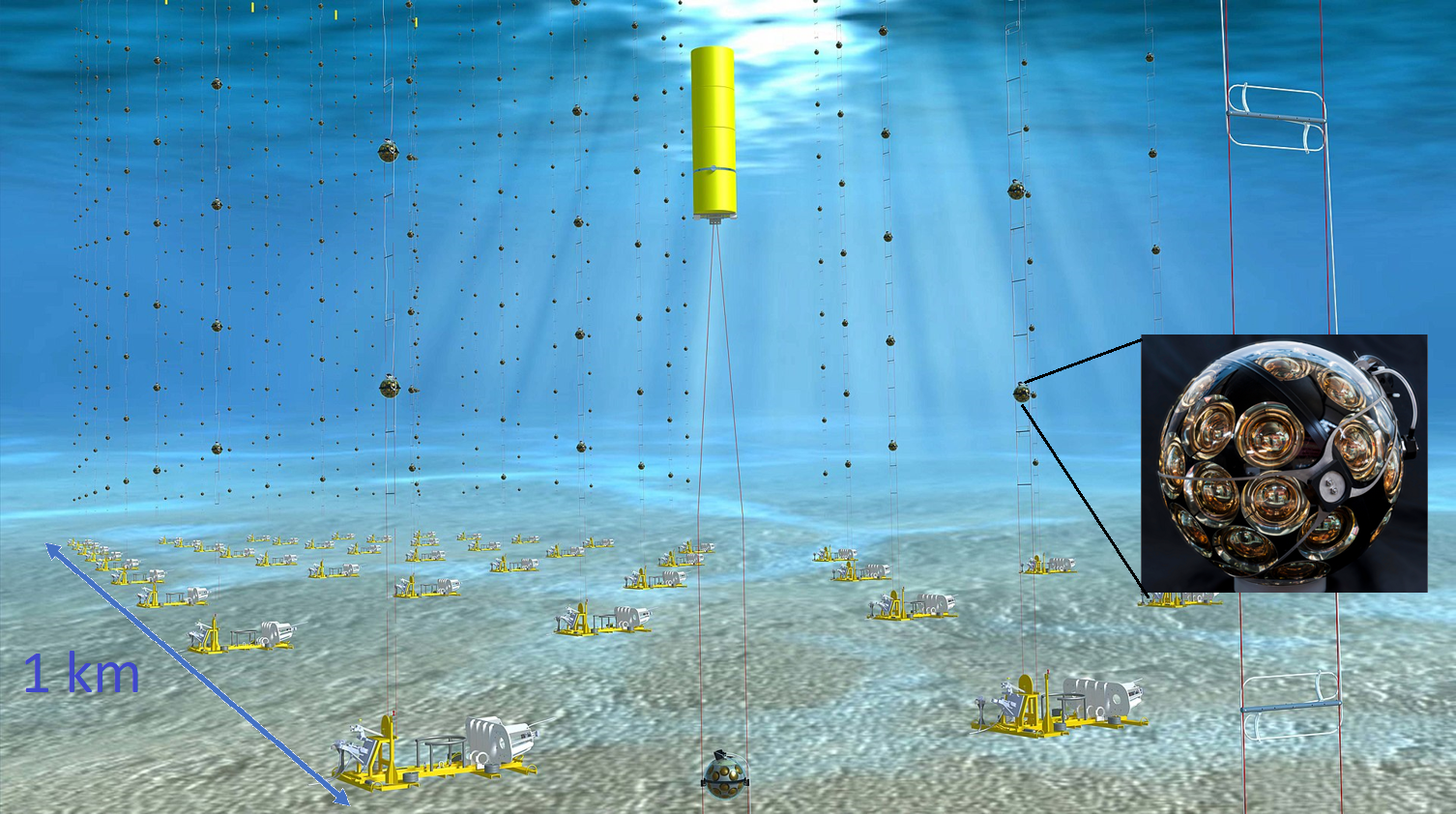}
\caption{Schematic view of a neutrino telescope. An array of photo sensors is embedded in a large volume of a transparent medium (water or ice) to determine the direction and energy of secondary charged particles induced by neutrino interactions using the Cherenkov radiation. The figure shows the layout of the KM3NeT detector, in construction in the Mediterranean Sea with one of the multi-PMT sensor (credit: KM3NeT Collaboration, km3net.org).}
\label{fig:KM3NeT}      
\end{figure*}

Neutrino telescopes consist of underwater/ice very large detectors ($\sim\! 1\,{\rm km}^3$ instrumented volume) able to observe secondary charged particles produced by neutrino interactions (Fig.~\ref{fig:KM3NeT}). 
Depending on neutrino flavour and interaction type, the event typologies can be either long tracks (for events induced by $\nu_\mu$ charged current (CC) interactions) or cascades (for $\nu_e$ and most $\nu_\tau$ CC interactions, and neutral current (NC) interactions). 
In the case of tracks, $2\pi$ sr of the sky are accessible, with an angular resolution down to 0.1$^\circ$-0.2$^\circ$ for the parent neutrino direction and a coarse energy estimation. 
For cascades, the angular resolution can be at most of 3$^\circ$-5$^\circ$ over the whole $4\pi$ of the sky, but with better energy determination. 
Atmospheric neutrinos represent an irreducible background for cosmic events: only at energies above tens of TeV, the number of astrophysical neutrinos overcomes that produced in the Earth atmosphere.
Below $\sim$100 TeV, cosmic neutrinos can be selected by a directional excess in a small solid angle, or in coincidence with electromagnetic or GW events.

Three detectors will be operational in the 2030s. The IceCube detector is currently operating at the South Pole, and it will be extended with its next-generation instrument, IceCube-Gen2 \cite{2020arXiv200804323T}. 
The KM3NeT arrays are under construction in the Mediterranean Sea \cite{KM3NeT_2016}, with a km$^3$ scale detector having an effective area larger than the present IceCube; KM3NeT will include at least two sites, one in Italy and one France. 
Finally, a third detector is the GVD neutrino telescope \cite{2019arXiv190805450B} under construction in the Baikal Lake (Russia), with larger sensitivity for neutrinos with $E> 100$ TeV.

Cosmic neutrinos are secondary particles produced through the interaction of protons (or nuclei) with matter or radiation fields, either close to the sources or during propagation in the interstellar/intergalactic medium. Neutrinos interact only through weak interactions and therefore escape dense astrophysical environments that are opaque to electromagnetic radiation. Unlike charged particles, neutrinos are not deflected by magnetic fields and point back to their source, thus settling the long-standing question of cosmic ray origin(s). 

Diffuse cosmic neutrinos have been observed with high statistical significance \cite{Aarsten2013}; however, their origin is still hotly debated. 
The neutrino detectors active in the 2030s will sharpen our understanding of the processes involved in Galactic and extragalactic objects at the highest energies, and their matter/radiation nearby environments. Their main goal will be to: 1) resolve the sources of high-energy neutrinos from the TeV to the EeV energies; 2) investigate cosmic particle acceleration processes particularly with multi-messenger observations; 3) study the extragalactic medium and understand the propagation of the highest energy particles in the Universe, by comparing neutrinos, X-rays and $\gamma$-rays  propagation.
Above $\sim$ 100 TeV (10$^{14}$ eV), the extragalactic space becomes opaque due to the scattering of $\gamma$-rays on the CMB and on the extragalactic background light. This leaves neutrinos as unique messengers to probe the most extreme particle accelerators in the cosmos. 

Neutrino telescopes operating in the 2030s will have sufficient effective area to identify some Galactic and extra-galactic steady sources. 
However, synergies with other prompt observations of transient events will significantly improve the discovery opportunities, as demonstrated by the recent observation of a neutrino event in spatial and temporal coincidence with an enhanced $\gamma$-ray emission state of the blazar TXS 0506+056 \cite{IC2018TXS}.
Different models foresee enhanced neutrino emission during flaring activities and hardness transition states in X-ray sources. Primary candidates are Galactic X-ray binaries, extragalactic flaring blazars, transient flares (GRBs) (see also \cite{Ciolfi2011}, this volume).
Neutrino telescopes can rapidly respond to an external alert. The presence of an event in the target sky region can be registered within 5-30 s. They can also provide fast alerts for interesting neutrino candidates of likely cosmic origin. All notable $\nu$ events from neutrino facilities will trigger alerts that will be distributed publicly to the community, with all the necessary metadata information.

For the above reasons, the emergent field of cosmic neutrino astrophysics will greatly benefit from coordinated operations with THESEUS. Multi-messenger observations can be used to identify blazars involved in p-$\gamma$ interactions that produce the neutrino emission. In particular, prompt X-ray observations will be the most sensitive probes for these types of correlated studies.
Another hadronic mechanism involved in the ultra-high energy cosmic rays acceleration is expected to be present in GRBs. The underlying physics that lies in the heart of GRB explosions is still unknown, and it has long been thought that in a significant fraction of explosions, the GRBs jets are choked. The expected outcome of a choked GRB jet may be neutrinos and X-rays \cite{Guetta2020}. Thus, detected neutrinos can be used as targets of opportunity for THESEUS to constrain the expected rate of choked GRBs. The X-ray cocoon emission is expected to be detected by the SXI instrument out to a distance of few hundreds of Mpc. THESEUS can therefore provide fundamental information about the rate of these events, as well as their temporal profile and physical parameters (such as energy released or stellar type, etc.), which in turn will improve our understanding of the GRB-supernovae connection. 

In the context of the international cooperation on multimessenger astronomy, THESEUS has the potential to become a crucial participant of future coordinated networks, such as the AMON (Astrophysical Multimessenger Observatory Network) \cite{AyalaSolares:2019iiy}, which currently includes many observational facilities across the globe. This will allow a full exploitation of data obtained by EM, neutrino and GW observatories by considering also subthreshold events.

\section{The Cherenkov Telescope Array (CTA)}
\label{sec:CTA}
The Cherenkov Telescope Array (CTA~\cite{CTA_book}) is a global, next-generation observatory studying very-high-energy (VHE) gamma rays in the energy range from tens of GeV to hundreds of TeV. The observatory will be implemented in two sites - Chile and the Canary Islands - and will comprise a range of Imaging Air Cherenkov Telescopes (IACTs) of different sizes. The current CTA design will allow us to obtain unprecedented sensitivity, as well as unique angular and energy resolutions over a large energy range. CTA is building on the success of the current generation of IACTs, namely H.E.S.S., MAGIC, VERITAS and FACT. In the first years after the completion of construction (expected around 2025), about half of the available observation time will be dedicated to Key Science Projects defined by the CTA Consortium. The fraction of the time allocated to guest observer programs will increase rapidly. It is thus expected that CTA will be a fully open observatory that coincides with the timescale of the THESEUS mission.\\

We expect strong synergies between the science programmes of CTA and THESEUS. One of the main science drivers for both observatories is the study of transient phenomena and especially GRBs. This field has recently seen tremendous breakthroughs with the detection of VHE emission from GRBs detected by H.E.S.S. (GRB~180720B~\cite{2019Natur.575..464A} and GRB~190829A) and MAGIC (GRB~190114C~\cite{2019Natur.575..455M,2019Natur.575..459M} and GRB~201216C). As an example, the detection of the close-by and very low luminosity burst GRB~190829A may indicate that the phase space of low luminosity events detectable in large numbers with THESEUS will enable many further joint CTA-THESEUS observations and studies. Access to this new phase space, combined with the high sensitivity of CTA and its fast reaction to multi-wavelength alerts in the 2030s over the full sky, will allow detailed joint studies of GRB light-curves across many orders of magnitude in energy and covering large time ranges. Moreover, THESEUS/IRT is best suited to follow-up these events that can explode in very dusty environments (e.g., GRBs 190114C and 190829A). The VHE gamma-ray horizon is limited by absorption on the extragalactic background radiation fields. Fortunately, the low CTA energy threshold will allow this horizon to be expanded  with respect to the one of the current generation of IACTs, and will thus benefit from the increased number of high-redshift GRBs detected by THESEUS. 

At lower redshifts, the joint CTA-THESEUS studies will naturally be extended into the multi-messenger domain through searches for counterparts to GW events. Given the large expected number of GW detections at the timescale of THESEUS, rapid and precise localisation of associated electromagnetic transients by THESEUS will lead to significantly increase the efficiency and performance of CTA observations of these high-priority targets. Gravitational waves encode important information (e.g. mass distribution, geometry, etc.) about the merging system even before it triggers the explosions detected as short GRBs in the electromagnetic domain. Combined observations of GWs, X-rays and VHE gamma-rays will thus help to significantly increase our understanding on how the progenitor system links to the observational appearance of the GRB. Significantly expanding on the first hints provided by the (non-)observations of GW170817~\cite{2017ApJ...850L..22A,2020ApJ...894L..16A}, we will be able for example to study the conditions of the burst progenitor, the parameters of the burst and the circumburst environment necessary to trigger VHE emission, the maximum energy reached and its possible scaling with the jet opening angle.

The study of short GRBs and their possible kilonova emission might be further explored by the connection of THESEUS with CTA. Recently, a hint of late-time TeV emission from the short GRB~160821B, for which an associated kilonova was found~\cite{2019MNRAS.489.2104T}, has been reported by MAGIC~\cite{2020arXiv201207193M}. The great opportunity provided by an instrument such as THESEUS to catch such infrared transients and the superior sensitivity of CTA with respect to current generation IACT telescopes will help to study the environment of the short GRBs and their possible associated kilonova explosions. 

\section{THESEUS as an observatory}
\label{sec:GOObs}

THESEUS is posed to provide a special opportunity for agile NIR and X-ray observations of a wide range of targets, from asteroids to the most distant AGN. Space-borne, sensitive NIR spectroscopy is a powerful capability, and wide-field sensitive X-ray monitoring can identify changes and priorities for follow-up studies. 

While in Survey Mode, the IRT, SXI and XGIS will be gathering data, with IRT pointed at a specific target. Hundreds of thousands of suitable targets are already known, and eROSITA, Euclid, Rubin and SKA will deepen and extend the range of relevant catalogs. 
A space-based infrared, and X-ray spectroscopic facility will be attractive to a wide range of investigators and address important questions in a plethora of scientific areas. While less powerful than JWST and Athena, the chance to use THESEUS to observe substantial samples of interesting sources, both known and newly-discovered, to appropriate depths and cadences, while the mission is searching for GRBs, provides opportunities for additional science. 
A user community interested in scales all the way from the Solar System to distant AGN can provide abundant desired targets for THESEUS as an observatory. 
THESEUS will maintain a list of core-programme targets, augmented with targets from a completed GO programme, with all observations planned and executed by the THESEUS mission operations team, while THESEUS operates in Survey Mode. The sensitivity of SXI and IRT are well matched and deliver images and spectra to useful depths in a fractional orbit spent staring at a specific target field. Existing catalogs of tens of thousands of targets will grow from the forthcoming very large wide-field Rubin optical (see sect.\ref{sec:LSST}) and eROSITA X-ray catalogs prior to launch. The numbers of known exoplanets continue to grow, and new generation of radio facilities are providing rich catalogs as precursors to the imaging of the SKA. 

\subsection{Key observatory science}
We briefly discuss the range and numbers of suitable targets for THESEUS as an observatory during the nominal mission, working out in cosmic distance. The co-alignment of the IRT with the part of the SXI FoV where the two units overlap ensures that the best X-ray spectra/limits will be obtained alongside every IR imaging/spectral target. 

A space-borne IR spectrograph is able to investigate a range of cometary emission and absorption features, without being restricted to specific atmospheric bands, and with full access to all water and ice features, impossible from the ground. Several tens of comets per year are likely to be observable as they pass through the inner solar system, evolving through their approach to and recession from perihelion. 

IR spectra of large samples of stars with transiting planets can be obtained by THESEUS. By 2030, tens of thousands of transiting planets will be known, spread widely over the sky, and with well-determined transit times, which can be scheduled well in advance to search for potential atmospheric signatures in IR absorption spectroscopy. IRT is more sensitive than the Atmospheric Remote-sensing Infrared Exoplanet Large-survey mission (ARIEL), and so carefully chosen extended planetary transit observations can be made for known targets, and for a substantial number of transiting planetary targets can be included in the observatory science target catalog. 

X-ray binaries and flaring stars can be discovered as bright X-ray and IR spectral targets by THESEUS, or highlighted by other observatories, and then confirmed and studied using THESEUS’s spectroscopic capabilities. Found predominantly in the Galactic Plane, many hundreds of bright events will occur during the nominal mission. 

A prompt spectroscopic IR survey for supernovae that are found taking place out to several 10s of Mpc, unencumbered by atmospheric effects, is likely to remain attractive beyond 2030, and will help to resolve remaining questions about the impact of environment and metallicity on the nature of supernovae and their reliability as standard candles. Without sensitive IR spectroscopy, these questions might not be resolved. 

The availability of the full spectral window is particularly helpful for observations of emission-line galaxies and AGN, for which key diagnostic lines are redshifted out of the optical band from the ground at redshifts $z\sim\!0.7$. Even ELTs cannot beat the atmosphere, and huge candidate samples will be cataloged over large fractions of the sky, color-selected from Rubin LSST surveys, in concert with the coverage of eROSITA, SKA and WISE. IRT will enable H$\alpha$ spectral surveys of interesting classes of the most luminous galaxies, and AGN all the way to $z\sim\! 2-3$. Furthermore, the THESEUS mission will provide a useful time baseline out to several years, to see potential changes in the appearance of AGN spectra, and to confirm any changes by revisiting selected examples.
While it will be impossible to include more than a few thousand galaxies and AGN in a spectral monitoring programme, the results of combining the wide-area data from LSST and WISE in the optical and IR, and with eROSITA in the X-ray, with the serendipitous wide-area coverage of SXI and XGIS, will allow new insight into the X-ray variability of large samples of AGN. 
There will be demand for tens of thousands of IRT spectral targets, from comets to distant AGN.  Given that several spectra can be obtained per orbit, a practical number of targets is of order 1000 each month, and tens of thousands of observations in parallel to Survey Mode during the nominal mission.

IRT observations will also be very valuable for in-depth studies of brown dwarfs, which represent the lowest-mass extension of the main-sequence in the Hertzsprung-Russell diagram, with luminosities and effective temperatures lower than those of M dwarfs. Their masses lie below the hydrogen-burning limit of $\sim$0.075 ~M$_{\odot}$, with radii of 1 R$_{\rm Jupiter}$ for mature objects, therefore their study can give important clues on star formation processes at the low mass end. The spectral resolution of IRT and its wavelength range of 0.8-1.6 $\mu$m are adequate to measuring  typical spectral features (specifically the FeH$_{Z}$, VO$_{Z}$, FeH$_{J}$ and KI$_{J}$ indices at 0.99, 1.06, 1.20, and 1.244, 1.253 $\mu$m), which are used to determine the brown dwarf spectral types (L, T, Y), and are excellent diagnostics of their youth \cite{allers13,Kirkpatrick99}.
IRT will provide high S/N spectra of brown dwarfs with R$\sim$400. This resolution is higher than that of the G141 grism on $Hubble$ WFC3, allowing "snap style" targets to be obtained and comparative studies, such as those of \cite{Manjavacas20}, to be performed at higher resolution than is currently possible. 
Rotation studies of brown dwarfs have also been performed with $Hubble$, creating spectral and spatial maps of their surfaces (e.g. \cite{Apai2013,Buenzli15,Yang2015} and in some cases (e.g. \cite{Yang2016}) have been performed simultaneously with observations taken at 3.6 and 4.5 $\mu$m with $Spitzer$. Following current launch schedules, THESEUS will be observing at the same time as ARIEL (launch 2029) and PLATO (launch 2026: extended mission phase), making it possible to obtain simultaneous near-IR spectra/photometry and optical photometry, or simultaneous near- and mid-IR spectra using the 1.95-7.80 $\mu$m AIRS instrument on ARIEL.

\subsection{Serendipitous detections}
THESEUS will operate with a very substantial field-of-view in a Survey Mode at a time when wide-field optical surveys with Rubin are mature, the eROSITA reference map of the X-ray sky will be available, and the SKA will be generating very deep radio images in the South. Many hundreds of thousands of interesting serendipitous sources will be detected using THESEUS’s SXI and XGIS instruments automatically during the mission, providing regular monitoring of a wide range of non-GRB transient sources for comparison against known high-energy sources. These classes of targets include not only the transients and variables discussed in \cite{Mereghetti2021}, but also interesting new targets from wide-field X-ray sky coverage, building on the eROSITA map of the sky. Furthermore, THESEUS’s sensitive IR/X-ray monitoring capability will provide a very useful tool for selecting targets for Athena and possible future large optical-NIR facilities.

One can also anticipate that the vast database of serendipitous sources (transients or not) collected by the THESEUS during its lifetime, when operating in survey mode with XGIS, SXI and IRT, or as part of IRT GO programs, complemented with other surveys  (e.g. SKA and Rubin/LSST), will also stimulate a large number of follow-up or monitoring programs with 1-4 m class telescopes, including robotic observatories (e.g. \cite{NRT}), thus further extending THESEUS' science impact on a wider international community. 

\begin{acknowledgements}

\noindent This work is partly supported by the AHEAD-2020 Project grant agreement 871158 of the European Union's Horizon 2020 Programme. Authors from INAF acknowledge support by ASI-INAF agreement n. 2018-29-HH.0. AG acknowledges financial support from the Slovenian Research Agency (grants P1-0031, I0-0033, J1-8136, J1-2460) and networking support by the COST Actions CA16104 GWverse and CA16214 PHAROS. MM is supported by the  Swiss National Science Foundation and  by the SwissMap National Center for Competence in Research. A.R. acknowledges support from the Premiale Supporto Arizona \& Italia. PTOB acknowledges support from the UK Space Agency. SDV and DP acknowledge support from CNES, focused on THESEUS.

\end{acknowledgements}

%
%


\bibliographystyle{spphys}       

\bibliography{All_refs}     

\begin{thebibliography}{10}
\providecommand{\url}[1]{{#1}}
\providecommand{\urlprefix}{URL }
\expandafter\ifx\csname urlstyle\endcsname\relax
  \providecommand{\doi}[1]{DOI \discretionary{}{}{}#1}\else
  \providecommand{\doi}{DOI \discretionary{}{}{}\begingroup
  \urlstyle{rm}\Url}\fi

\bibitem{LVC-BNS}
B.P. {Abbott}, R.~{Abbott}, T.D. {Abbott}, F.~{Acernese}, K.~{Ackley},
  C.~{Adams}, T.~{Adams}, P.~{Addesso}, R.X. {Adhikari}, V.B. {Adya}, et~al.,
  Phys. Rev. Lett. \textbf{119}(16), 161101 (2017).
\newblock \doi{10.1103/PhysRevLett.119.161101}

\bibitem{Amati2021}
{L. Amati et al.}, submitted to Experimental Astronomy  (2021)

\bibitem{Mereghetti2021}
{S. Mereghetti et al.}, submitted to Experimental Astronomy  (2021)

\bibitem{Ciolfi2021}
{R. Ciolfi et al.}, submitted to Experimental Astronomy  (2021)

\bibitem{Tanvir2021}
{N.R. Tanvir et al.}, submitted to Experimental Astronomy  (2021)

\bibitem{Ghirlanda2021}
{G. Ghirlanda et al.}, submitted to Experimental Astronomy  (2021)

\bibitem{Walsh2020}
S.~{Walsh}, S.~{McBreen}, A.~{Martin-Carrillo}, T.~{Dauser}, N.~{Wijers},
  J.~{Wilms}, J.~{Schaye}, D.~{Barret}, \aap \textbf{642}, A24 (2020).
\newblock \doi{10.1051/0004-6361/202037775}

\bibitem{Hartoog2015}
O.E. {Hartoog}, D.~{Malesani}, J.P.U. {Fynbo}, T.~{Goto}, T.~{Kr{\"u}hler},
  P.M. {Vreeswijk}, A.~{De Cia}, D.~{Xu}, P.~{M{\o}ller}, S.~{Covino},
  V.~{D'Elia}, H.~{Flores}, P.~{Goldoni}, J.~{Hjorth}, P.~{Jakobsson}, J.K.
  {Krogager}, L.~{Kaper}, C.~{Ledoux}, A.J. {Levan}, B.~{Milvang-Jensen},
  J.~{Sollerman}, M.~{Sparre}, G.~{Tagliaferri}, N.R. {Tanvir}, A.~{de Ugarte
  Postigo}, S.D. {Vergani}, K.~{Wiersema}, J.~{Datson}, R.~{Salinas},
  K.~{Mikkelsen}, N.~{Aghanim}, \aap \textbf{580}, A139 (2015).
\newblock \doi{10.1051/0004-6361/201425001}

\bibitem{Pian2017}
E.~{Pian}, P.~{D'Avanzo}, S.~{Benetti}, M.~{Branchesi}, E.~{Brocato},
  S.~{Campana}, E.~{Cappellaro}, S.~{Covino}, V.~{D'Elia}, J.P.U. {Fynbo},
  F.~{Getman}, G.~{Ghirlanda}, G.~{Ghisellini}, A.~{Grado}, G.~{Greco},
  J.~{Hjorth}, C.~{Kouveliotou}, A.~{Levan}, L.~{Limatola}, D.~{Malesani}, P.A.
  {Mazzali}, A.~{Melandri}, P.~{M{\o}ller}, L.~{Nicastro}, E.~{Palazzi},
  S.~{Piranomonte}, A.~{Rossi}, O.S. {Salafia}, J.~{Selsing}, G.~{Stratta},
  M.~{Tanaka}, N.R. {Tanvir}, L.~{Tomasella}, D.~{Watson}, S.~{Yang},
  L.~{Amati}, L.A. {Antonelli}, S.~{Ascenzi}, M.G. {Bernardini}, M.~{Bo{\"e}r},
  F.~{Bufano}, A.~{Bulgarelli}, M.~{Capaccioli}, P.~{Casella}, A.J.
  {Castro-Tirado}, E.~{Chassande-Mottin}, R.~{Ciolfi}, C.M. {Copperwheat},
  M.~{Dadina}, G.~{De Cesare}, A.~{di Paola}, Y.Z. {Fan}, B.~{Gendre},
  G.~{Giuffrida}, A.~{Giunta}, L.K. {Hunt}, G.L. {Israel}, Z.P. {Jin}, M.M.
  {Kasliwal}, S.~{Klose}, M.~{Lisi}, F.~{Longo}, E.~{Maiorano}, M.~{Mapelli},
  N.~{Masetti}, L.~{Nava}, B.~{Patricelli}, D.~{Perley}, A.~{Pescalli},
  T.~{Piran}, A.~{Possenti}, L.~{Pulone}, M.~{Razzano}, R.~{Salvaterra},
  P.~{Schipani}, M.~{Spera}, A.~{Stamerra}, L.~{Stella}, G.~{Tagliaferri},
  V.~{Testa}, E.~{Troja}, M.~{Turatto}, S.D. {Vergani}, D.~{Vergani}, Nature
  \textbf{551}(7678), 67 (2017).
\newblock \doi{10.1038/nature24298}

\bibitem{McGuire2016}
J.T.W. {McGuire}, N.R. {Tanvir}, A.J. {Levan}, M.~{Trenti}, E.R. {Stanway},
  J.M. {Shull}, K.~{Wiersema}, D.A. {Perley}, R.L.C. {Starling}, M.~{Bremer},
  J.T. {Stocke}, J.~{Hjorth}, J.E. {Rhoads}, E.~{Curtis-Lake}, S.~{Schulze},
  E.M. {Levesque}, B.~{Robertson}, J.P.U. {Fynbo}, R.S. {Ellis}, A.S.
  {Fruchter}, \apj \textbf{825}(2), 135 (2016).
\newblock \doi{10.3847/0004-637X/825/2/135}

\bibitem{Ghirlanda2015}
G.~{Ghirlanda}, R.~{Salvaterra}, G.~{Ghisellini}, S.~{Mereghetti},
  G.~{Tagliaferri}, S.~{Campana}, J.P. {Osborne}, P.~{O'Brien}, N.~{Tanvir},
  D.~{Willingale}, L.~{Amati}, S.~{Basa}, M.G. {Bernardini}, D.~{Burlon},
  S.~{Covino}, P.~{D'Avanzo}, F.~{Frontera}, D.~{G{\"o}tz}, A.~{Melandri},
  L.~{Nava}, L.~{Piro}, S.D. {Vergani}, \mnras \textbf{448}(3), 2514 (2015).
\newblock \doi{10.1093/mnras/stv183}

\bibitem{Rossi2020}
A.~{Rossi}, G.~{Stratta}, E.~{Maiorano}, D.~{Spighi}, N.~{Masetti},
  E.~{Palazzi}, A.~{Gardini}, A.~{Melandri}, L.~{Nicastro}, E.~{Pian},
  M.~{Branchesi}, M.~{Dadina}, V.~{Testa}, E.~{Brocato}, S.~{Benetti},
  R.~{Ciolfi}, S.~{Covino}, V.~{D'Elia}, A.~{Grado}, L.~{Izzo}, A.~{Perego},
  S.~{Piranomonte}, R.~{Salvaterra}, J.~{Selsing}, L.~{Tomasella}, S.~{Yang},
  D.~{Vergani}, L.~{Amati}, J.B. {Stephen}, Mon. Not. R. Astron. Soc.
  \textbf{493}(3), 3379 (2020).
\newblock \doi{10.1093/mnras/staa479}

\bibitem{Maiorano2018}
E.~{Maiorano}, L.~{Amati}, A.~{Rossi}, G.~{Stratta}, E.~{Palazzi},
  L.~{Nicastro}, \memsai \textbf{89}, 181 (2018)

\bibitem{MAORY}
G.~{Fiorentino}, M.~{Bellazzini}, P.~{Ciliegi}, G.~{Chauvin}, S.~{Dout{\'e}},
  V.~{D'Orazi}, E.~{Maiorano}, F.~{Mannucci}, M.~{Mapelli}, L.~{Podio},
  P.~{Saracco}, M.~{Spavone}, arXiv e-prints arXiv:1712.04222 (2017)

\bibitem{SKA2013}
M.~{Huynh}, J.~{Lazio}, arXiv e-prints arXiv:1311.4288 (2013)

\bibitem{Mitra_Reion}
S.~{Mitra}, T.R. {Choudhury}, A.~{Ferrara}, Mon. Not. R. Astron. Soc.
  \textbf{473}(1), 1416 (2018).
\newblock \doi{10.1093/mnras/stx2443}

\bibitem{Xu2011}
Y.~{Xu}, A.~{Ferrara}, X.~{Chen}, Mon. Not. R. Astron. Soc. \textbf{410}(3),
  2025 (2011).
\newblock \doi{10.1111/j.1365-2966.2010.17579.x}

\bibitem{Gehrels2004-Swift}
N.~{Gehrels}, G.~{Chincarini}, P.~{Giommi}, K.O. {Mason}, J.A. {Nousek}, A.A.
  {Wells}, N.E. {White}, S.D. {Barthelmy}, D.N. {Burrows}, L.R. {Cominsky},
  et~al., Astrophys. J. \textbf{611}, 1005 (2004).
\newblock \doi{10.1086/422091}

\bibitem{Totani06}
T.~{Totani}, N.~{Kawai}, G.~{Kosugi}, K.~{Aoki}, T.~{Yamada}, M.~{Iye},
  K.~{Ohta}, T.~{Hattori}, \pasj \textbf{58}(3), 485 (2006).
\newblock \doi{10.1093/pasj/58.3.485}

\bibitem{Gallerani08}
S.~{Gallerani}, R.~{Salvaterra}, A.~{Ferrara}, T.R. {Choudhury}, Mon. Not. R.
  Astron. Soc. \textbf{388}(1), L84 (2008).
\newblock \doi{10.1111/j.1745-3933.2008.00504.x}

\bibitem{Ghirlanda2013}
G.~{Ghirlanda}, R.~{Salvaterra}, D.~{Burlon}, S.~{Campana}, A.~{Melandri}, M.G.
  {Bernardini}, S.~{Covino}, P.~{D'Avanzo}, V.~{D'Elia}, G.~{Ghisellini},
  L.~{Nava}, I.~{Prandoni}, L.~{Sironi}, G.~{Tagliaferri}, S.D. {Vergani},
  A.~{Wolter}, \mnras \textbf{435}(3), 2543 (2013).
\newblock \doi{10.1093/mnras/stt1466}

\bibitem{Perley2014}
D.A. {Perley}, S.B. {Cenko}, A.~{Corsi}, N.R. {Tanvir}, A.J. {Levan}, D.A.
  {Kann}, E.~{Sonbas}, K.~{Wiersema}, W.~{Zheng}, X.H. {Zhao}, J.M. {Bai},
  M.~{Bremer}, A.J. {Castro-Tirado}, L.~{Chang}, K.I. {Clubb}, D.~{Frail},
  A.~{Fruchter}, E.~{G{\"o}{\u{g}}{\"u}{\c{s}}}, J.~{Greiner}, T.~{G{\"u}ver},
  A.~{Horesh}, A.V. {Filippenko}, S.~{Klose}, J.~{Mao}, A.N. {Morgan}, A.S.
  {Pozanenko}, S.~{Schmidl}, B.~{Stecklum}, M.~{Tanga}, A.A. {Volnova}, A.E.
  {Volvach}, J.G. {Wang}, J.M. {Winters}, Y.X. {Xin}, \apj \textbf{781}(1), 37
  (2014).
\newblock \doi{10.1088/0004-637X/781/1/37}

\bibitem{Laskar18}
T.~{Laskar}, K.D. {Alexander}, E.~{Berger}, C.~{Guidorzi}, R.~{Margutti}, W.f.
  {Fong}, C.D. {Kilpatrick}, P.~{Milne}, M.R. {Drout}, C.G. {Mundell},
  S.~{Kobayashi}, R.~{Lunnan}, R.~{Barniol Duran}, K.M. {Menten}, K.~{Ioka},
  P.K.G. {Williams}, \apj \textbf{862}(2), 94 (2018).
\newblock \doi{10.3847/1538-4357/aacbcc}

\bibitem{Fernandez2021}
J.J. {Fern{\'a}ndez}, S.~{Kobayashi}, G.P. {Lamb}, arXiv e-prints
  arXiv:2101.05138 (2021)

\bibitem{Mooley2018}
K.P. {Mooley}, A.T. {Deller}, O.~{Gottlieb}, E.~{Nakar}, G.~{Hallinan},
  S.~{Bourke}, D.A. {Frail}, A.~{Horesh}, A.~{Corsi}, K.~{Hotokezaka}, \nat
  \textbf{561}(7723), 355 (2018).
\newblock \doi{10.1038/s41586-018-0486-3}

\bibitem{Ghirlanda2019}
G.~{Ghirlanda}, O.S. {Salafia}, Z.~{Paragi}, M.~{Giroletti}, J.~{Yang},
  B.~{Marcote}, J.~{Blanchard}, I.~{Agudo}, T.~{An}, M.G. {Bernardini},
  R.~{Beswick}, M.~{Branchesi}, S.~{Campana}, C.~{Casadio}, E.~{Chassand
  e-Mottin}, M.~{Colpi}, S.~{Covino}, P.~{D'Avanzo}, V.~{D'Elia}, S.~{Frey},
  M.~{Gawronski}, G.~{Ghisellini}, L.I. {Gurvits}, P.G. {Jonker}, H.J. {van
  Langevelde}, A.~{Melandri}, J.~{Moldon}, L.~{Nava}, A.~{Perego}, M.A.
  {Perez-Torres}, C.~{Reynolds}, R.~{Salvaterra}, G.~{Tagliaferri},
  T.~{Venturi}, S.D. {Vergani}, M.~{Zhang}, Science \textbf{363}(6430), 968
  (2019).
\newblock \doi{10.1126/science.aau8815}

\bibitem{2015A&A...578A..71G}
G.~{Ghirlanda}, R.~{Salvaterra}, S.~{Campana}, S.D. {Vergani}, J.~{Japelj},
  M.G. {Bernardini}, D.~{Burlon}, P.~{D'Avanzo}, A.~{Melandri}, A.~{Gomboc},
  F.~{Nappo}, R.~{Paladini}, A.~{Pescalli}, O.S. {Salafia}, G.~{Tagliaferri},
  \aap \textbf{578}, A71 (2015).
\newblock \doi{10.1051/0004-6361/201526112}

\bibitem{Marcote2019}
B.~{Marcote}, K.~{Nimmo}, O.S. {Salafia}, Z.~{Paragi}, J.W.T. {Hessels},
  E.~{Petroff}, R.~{Karuppusamy}, \apjl \textbf{876}(1), L14 (2019).
\newblock \doi{10.3847/2041-8213/ab1aad}

\bibitem{Shivvers2011}
I.~{Shivvers}, E.~{Berger}, \apj \textbf{734}(1), 58 (2011).
\newblock \doi{10.1088/0004-637X/734/1/58}

\bibitem{Burlon2016}
D.~{Burlon}, T.~{Murphy}, G.~{Ghirlanda}, P.J. {Hancock}, R.~{Parry},
  R.~{Salvaterra}, \mnras \textbf{459}(3), 3356 (2016).
\newblock \doi{10.1093/mnras/stw905}

\bibitem{Murphy+2018}
E.J. {Murphy}, A.~{Bolatto}, S.~{Chatterjee}, C.M. {Casey}, L.~{Chomiuk},
  D.~{Dale}, I.~{de Pater}, M.~{Dickinson}, J.D. {Francesco}, G.~{Hallinan},
  A.~{Isella}, K.~{Kohno}, S.R. {Kulkarni}, C.~{Lang}, T.J.W. {Lazio}, A.K.
  {Leroy}, L.~{Loinard}, T.J. {Maccarone}, B.C. {Matthews}, R.A. {Osten}, M.J.
  {Reid}, D.~{Riechers}, N.~{Sakai}, F.~{Walter}, D.~{Wilner}, in \emph{Science
  with a Next Generation Very Large Array}, \emph{Astronomical Society of the
  Pacific Conference Series}, vol. 517, ed. by E.~{Murphy} (2018),
  \emph{Astronomical Society of the Pacific Conference Series}, vol. 517, p.~3

\bibitem{Selina+2018}
R.J. {Selina}, E.J. {Murphy}, M.~{McKinnon}, A.~{Beasley}, B.~{Butler},
  C.~{Carilli}, B.~{Clark}, S.~{Durand}, A.~{Erickson}, W.~{Grammer},
  R.~{Hiriart}, J.~{Jackson}, B.~{Kent}, B.~{Mason}, M.~{Morgan}, O.Y. {Ojeda},
  V.~{Rosero}, W.~{Shillue}, S.~{Sturgis}, D.~{Urbain}, in \emph{Science with a
  Next Generation Very Large Array}, \emph{Astronomical Society of the Pacific
  Conference Series}, vol. 517, ed. by E.~{Murphy} (2018), \emph{Astronomical
  Society of the Pacific Conference Series}, vol. 517, p.~15

\bibitem{LR2018}
N.~{Lloyd-Ronning}, in \emph{Science with a Next Generation Very Large Array},
  \emph{Astronomical Society of the Pacific Conference Series}, vol. 517, ed.
  by E.~{Murphy} (2018), \emph{Astronomical Society of the Pacific Conference
  Series}, vol. 517, p. 701

\bibitem{vanVelzen+2018}
S.~{van Velzen}, G.C. {Bower}, B.D. {Metzger}, in \emph{Science with a Next
  Generation Very Large Array}, \emph{Astronomical Society of the Pacific
  Conference Series}, vol. 517, ed. by E.~{Murphy} (2018), \emph{Astronomical
  Society of the Pacific Conference Series}, vol. 517, p. 737

\bibitem{Lister+2018}
M.L. {Lister}, K.I. {Kellermann}, P.~{Kharb}, in \emph{Science with a Next
  Generation Very Large Array}, \emph{Astronomical Society of the Pacific
  Conference Series}, vol. 517, ed. by E.~{Murphy} (2018), \emph{Astronomical
  Society of the Pacific Conference Series}, vol. 517, p. 619

\bibitem{LSST_book}
{P.A. Abell et al. (LSST Science Collaboration)}.
\newblock {LSST Science Book, Version 2.0} (2009)

\bibitem{2014ApJ...796...53R}
S.T. {Ridgway}, T.~{Matheson}, K.J. {Mighell}, K.A. {Olsen}, S.B. {Howell},
  \apj \textbf{796}(1), 53 (2014).
\newblock \doi{10.1088/0004-637X/796/1/53}

\bibitem{LSSTAlerts}
{Graham, M.L., Bellm, E., Guy, L., Slater, C. T., Dubois-Felsmann, G. and the
  Data Management System Science Team}.
\newblock {LSST Data Management - LSST Alerts: Key Numbers} (2020)

\bibitem{2020ApJ...890...73B}
K.~{Bricman}, A.~{Gomboc}, \apj \textbf{890}(1), 73 (2020).
\newblock \doi{10.3847/1538-4357/ab6989}

\bibitem{LSSTObsStrategy}
P.~{Marshall}, W.~{Clarkson}, O.~{Shemmer}, R.~{Biswas}, M.~{de Val-Borro},
  J.~{Rho}, L.~{Jones}, T.~{Anguita}, S.~{Ridgway}, F.~{Bianco}, Z.~{Ivezic},
  M.~{Lochner}, J.~{Meyers}, K.~{Vivas}, M.~{Graham}, C.~{Claver}, S.~{Digel},
  V.~{Kasliwal}, P.M. {McGehee}, E.~{Gawiser}, E.~{Bellm}, L.~{Walkowicz},
  K.~{Olsen}, P.~{Yoachim}, K.~{Bell}, D.~{Nidever}, M.~{Lund}, A.~{Connolly},
  I.~{Arcavi}, H.~{Awan}.
\newblock {Lsst Science Collaborations Observing Strategy White Paper:
  ``Science-Driven Optimization Of The Lsst Observing Strategy''} (2017).
\newblock \doi{10.5281/zenodo.842713}

\bibitem{2010PhDT.......241S}
M.~{Solontoi}, {Comets in Large Sky Surveys: From SDSS to LSST}.
\newblock Ph.D. thesis, University of Washington (2010)

\bibitem{Punturo:2010zz}
M.~Punturo, et~al., Class. Quant. Grav. \textbf{27}, 194002 (2010).
\newblock \doi{10.1088/0264-9381/27/19/194002}

\bibitem{2020JCAP...03..050M}
M.~{Maggiore}, C.~{Van Den Broeck}, N.~{Bartolo}, E.~{Belgacem}, D.~{Bertacca},
  M.A. {Bizouard}, M.~{Branchesi}, S.~{Clesse}, S.~{Foffa},
  J.~{Garc{\'\i}a-Bellido}, S.~{Grimm}, J.~{Harms}, T.~{Hinderer},
  S.~{Matarrese}, C.~{Palomba}, M.~{Peloso}, A.~{Ricciardone},
  M.~{Sakellariadou}, \jcap \textbf{2020}(3), 050 (2020).
\newblock \doi{10.1088/1475-7516/2020/03/050}

\bibitem{Evans:2016mbw}
B.P. Abbott, et~al., Class. Quant. Grav. \textbf{34}(4), 044001 (2017).
\newblock \doi{10.1088/1361-6382/aa51f4}

\bibitem{Reitze:2019iox}
D.~Reitze, et~al., Bull. Am. Astron. Soc. \textbf{51}(7), 035 (2019)

\bibitem{GWIC2020}
M.~{Bailes}, B.K. {Berger}, P.R. {Brady}, M.~{Branchesi}, K.~{Danzmann},
  M.~{Evans}, K.~{Holley-Bockelmann}, B.R. {Iyer}, T.~{Kajita},
  S.~{Katsanevas}, M.~{Kramer}, A.~{Lazzarini}, L.~{Lehner}, G.~{Losurdo},
  H.~{Lück}, D.E. {McClelland}, M.A. {McLaughlin}, M.~{Punturo}, S.~{Ransom},
  S.~{Raychaudhury}, D.H. {Reitze}, F.~{Ricci}, S.~{Rowan}, Y.~{Saito}, G.H.
  {Sanders}, B.S. {Sathyaprakash}, B.F. {Schutz}, A.~{Sesana}, H.~{Shinkai},
  X.~{Siemens}, D.H. {Shoemaker}, J.~{Thorpe}, J.F.J. {van den Brand},
  S.~{Vitale}, Nature Reviews Physics  (2021).
\newblock \doi{10.1038/s42254-021-00303-8}

\bibitem{LVK-LRR-2020-short}
{B.P. Abbott et al. (Kagra Collaboration, Ligo Scientific Collaboration, and
  VIRGO Collaboration)}, Living Reviews in Relativity \textbf{23}(1), 3 (2020).
\newblock \doi{10.1007/s41114-020-00026-9}

\bibitem{Adhikari2020}
{R.X.~Adhikari et al.}, Classical and Quantum Gravity \textbf{37}(16), 165003
  (2020).
\newblock \doi{10.1088/1361-6382/ab9143}

\bibitem{Ascenzi2020}
S.~{Ascenzi}, G.~{Oganesyan}, O.S. {Salafia}, M.~{Branchesi}, G.~{Ghirlanda},
  S.~{Dall'Osso}, Astron. Astrophys. \textbf{641}, A61 (2020).
\newblock \doi{10.1051/0004-6361/202038265}

\bibitem{ETsensitivity}
S.~{Hild}, M.~{Abernathy}, F.~{Acernese}, P.~{Amaro-Seoane}, N.~{Andersson},
  K.~{Arun}, F.~{Barone}, B.~{Barr}, M.~{Barsuglia}, e.a. {Beker}, Classical
  and Quantum Gravity \textbf{28}(9), 094013 (2011).
\newblock \doi{10.1088/0264-9381/28/9/094013}

\bibitem{2021MNRAS.502.4877S}
F.~{Santoliquido}, M.~{Mapelli}, N.~{Giacobbo}, Y.~{Bouffanais}, M.C. {Artale},
  \mnras \textbf{502}(4), 4877 (2021).
\newblock \doi{10.1093/mnras/stab280}

\bibitem{PhysRevD.97.104064}
C.~Mills, V.~Tiwari, S.~Fairhurst, Phys. Rev. D \textbf{97}, 104064 (2018).
\newblock \doi{10.1103/PhysRevD.97.104064}.
\newblock \urlprefix\url{https://link.aps.org/doi/10.1103/PhysRevD.97.104064}

\bibitem{PhysRevLett.94.201101}
M.~Shibata, Phys. Rev. Lett. \textbf{94}, 201101 (2005).
\newblock \doi{10.1103/PhysRevLett.94.201101}.
\newblock
  \urlprefix\url{https://link.aps.org/doi/10.1103/PhysRevLett.94.201101}

\bibitem{Stergioulas2011}
N.~{Stergioulas}, A.~{Bauswein}, K.~{Zagkouris}, H.T. {Janka}, Mon. Not. R.
  Astron. Soc \textbf{418}, 427 (2011).
\newblock \doi{10.1111/j.1365-2966.2011.19493.x}

\bibitem{2015PhRvD..91l4056B}
A.~{Bauswein}, N.~{Stergioulas}, Phys. Rev. D \textbf{91}(12), 124056 (2015).
\newblock \doi{10.1103/PhysRevD.91.124056}

\bibitem{2017ApJ...851L..16A}
B.P. {Abbott}, R.~{Abbott}, T.D. {Abbott}, F.~{Acernese}, K.~{Ackley},
  C.~{Adams}, T.~{Adams}, P.~{Addesso}, R.X. {Adhikari}, V.B. {Adya}, et~al.,
  Astrophys. J. Lett. \textbf{851}(1), L16 (2017).
\newblock \doi{10.3847/2041-8213/aa9a35}

\bibitem{2020ApJ...892L...3A}
B.P. {Abbott}, R.~{Abbott}, T.D. {Abbott}, S.~{Abraham}, F.~{Acernese},
  K.~{Ackley}, C.~{Adams}, R.X. {Adhikari}, V.B. {Adya}, C.~{Affeldt}, et~al.,
  Astrophys. J. Lett. \textbf{892}(1), L3 (2020).
\newblock \doi{10.3847/2041-8213/ab75f5}

\bibitem{2014PhRvD..90f2004C}
J.~{Clark}, A.~{Bauswein}, L.~{Cadonati}, H.T. {Janka}, C.~{Pankow},
  N.~{Stergioulas}, Phys. Rev. D \textbf{90}(6), 062004 (2014).
\newblock \doi{10.1103/PhysRevD.90.062004}

\bibitem{2019PhRvD.100j4029B}
M.~{Breschi}, S.~{Bernuzzi}, F.~{Zappa}, M.~{Agathos}, A.~{Perego},
  D.~{Radice}, A.~{Nagar}, Phys. Rev. D \textbf{100}(10), 104029 (2019).
\newblock \doi{10.1103/PhysRevD.100.104029}

\bibitem{2019PhRvD..99d4014T}
A.~{Torres-Rivas}, K.~{Chatziioannou}, A.~{Bauswein}, J.A. {Clark}, Phys. Rev.
  D \textbf{99}(4), 044014 (2019).
\newblock \doi{10.1103/PhysRevD.99.044014}

\bibitem{2020PhRvD.102d3011E}
P.J. {Easter}, S.~{Ghonge}, P.D. {Lasky}, A.R. {Casey}, J.A. {Clark},
  F.~{Hernandez Vivanco}, K.~{Chatziioannou}, Phys. Rev. D \textbf{102}(4),
  043011 (2020).
\newblock \doi{10.1103/PhysRevD.102.043011}

\bibitem{2020PhRvL.125z1101H}
C.J. {Haster}, K.~{Chatziioannou}, A.~{Bauswein}, J.A. {Clark}, Phys. Rev.
  Lett. \textbf{125}(26), 261101 (2020).
\newblock \doi{10.1103/PhysRevLett.125.261101}

\bibitem{Margalit2019}
B.~{Margalit}, B.D. {Metzger}, Astropys. J. Lett. \textbf{880}(1), L15 (2019).
\newblock \doi{10.3847/2041-8213/ab2ae2}

\bibitem{2020IJMPD..2941015F}
J.L. {Friedman}, N.~{Stergioulas}, International Journal of Modern Physics D
  \textbf{29}(11), 2041015-632 (2020).
\newblock \doi{10.1142/S0218271820410151}

\bibitem{2020PhRvL.125n1103B}
A.~{Bauswein}, S.~{Blacker}, V.~{Vijayan}, N.~{Stergioulas},
  K.~{Chatziioannou}, J.A. {Clark}, N.U.F. {Bastian}, D.B. {Blaschke},
  M.~{Cierniak}, T.~{Fischer}, Phys. Rev. Lett. \textbf{125}(14), 141103
  (2020).
\newblock \doi{10.1103/PhysRevLett.125.141103}

\bibitem{2020arXiv201004461B}
A.~{Bauswein}, S.~{Blacker}, G.~{Lioutas}, T.~{Soultanis}, V.~{Vijayan},
  N.~{Stergioulas}, arXiv e-prints arXiv:2010.04461 (2020)

\bibitem{Maggiore2020}
M.~{Maggiore}, C.~{Van Den Broeck}, N.~{Bartolo}, E.~{Belgacem}, D.~{Bertacca},
  M.A. {Bizouard}, M.~{Branchesi}, S.~{Clesse}, S.~{Foffa},
  J.~{Garc{\'\i}a-Bellido}, S.~{Grimm}, J.~{Harms}, T.~{Hinderer},
  S.~{Matarrese}, C.~{Palomba}, M.~{Peloso}, A.~{Ricciardone},
  M.~{Sakellariadou}, J. Cosmol. Astropart. Phys. \textbf{2020}(3), 050 (2020).
\newblock \doi{10.1088/1475-7516/2020/03/050}

\bibitem{Belgacem:2018lbp}
E.~Belgacem, Y.~Dirian, S.~Foffa, M.~Maggiore, Phys. Rev. \textbf{D98}, 023510
  (2018).
\newblock \doi{10.1103/PhysRevD.98.023510}

\bibitem{Belgacem:2019lwx}
E.~Belgacem, Y.~Dirian, A.~Finke, S.~Foffa, M.~Maggiore, JCAP \textbf{1911},
  022 (2019).
\newblock \doi{10.1088/1475-7516/2019/11/022}

\bibitem{PBH2020}
V.~{De Luca}, G.~{Franciolini}, P.~{Pani}, A.~{Riotto}, Physical Review D
  \textbf{102}(4), 043505 (2020).
\newblock \doi{10.1103/PhysRevD.102.043505}

\bibitem{LISA}
P.~{Amaro-Seoane}, H.~{Audley}, S.~{Babak}, J.~{Baker}, E.~{Barausse},
  P.~{Bender}, E.~{Berti}, P.~{Binetruy}, M.~{Born}, D.~{Bortoluzzi},
  J.~{Camp}, C.~{Caprini}, V.~{Cardoso}, M.~{Colpi}, J.~{Conklin},
  N.~{Cornish}, C.~{Cutler}, K.~{Danzmann}, R.~{Dolesi}, L.~{Ferraioli},
  V.~{Ferroni}, E.~{Fitzsimons}, J.~{Gair}, L.~{Gesa Bote}, D.~{Giardini},
  F.~{Gibert}, C.~{Grimani}, H.~{Halloin}, G.~{Heinzel}, T.~{Hertog},
  M.~{Hewitson}, K.~{Holley-Bockelmann}, D.~{Hollington}, M.~{Hueller},
  H.~{Inchauspe}, P.~{Jetzer}, N.~{Karnesis}, C.~{Killow}, A.~{Klein},
  B.~{Klipstein}, N.~{Korsakova}, S.L. {Larson}, J.~{Livas}, I.~{Lloro},
  N.~{Man}, D.~{Mance}, J.~{Martino}, I.~{Mateos}, K.~{McKenzie}, S.T.
  {McWilliams}, C.~{Miller}, G.~{Mueller}, G.~{Nardini}, G.~{Nelemans},
  M.~{Nofrarias}, A.~{Petiteau}, P.~{Pivato}, E.~{Plagnol}, E.~{Porter},
  J.~{Reiche}, D.~{Robertson}, N.~{Robertson}, E.~{Rossi}, G.~{Russano},
  B.~{Schutz}, A.~{Sesana}, D.~{Shoemaker}, J.~{Slutsky}, C.F. {Sopuerta},
  T.~{Sumner}, N.~{Tamanini}, I.~{Thorpe}, M.~{Troebs}, M.~{Vallisneri},
  A.~{Vecchio}, D.~{Vetrugno}, S.~{Vitale}, M.~{Volonteri}, G.~{Wanner},
  H.~{Ward}, P.~{Wass}, W.~{Weber}, J.~{Ziemer}, P.~{Zweifel}, arXiv e-prints
  arXiv:1702.00786 (2017)

\bibitem{Aarsten2013}
{IceCube Collaboration}, Science \textbf{342}(6161), 1242856 (2013).
\newblock \doi{10.1126/science.1242856}

\bibitem{2020arXiv200804323T}
{The IceCube-Gen2 Collaboration}, {:}, M.G. {Aartsen}, R.~{Abbasi},
  M.~{Ackermann}, e.a. {Adams}, arXiv e-prints arXiv:2008.04323 (2020)

\bibitem{KM3NeT_2016}
S.~Adrián-Martínez, M.~Ageron, F.~Aharonian, S.~Aiello, A.~Albert, F.~Ameli,
  E.~Anassontzis, M.~Andre, G.~Androulakis, M.~Anghinolfi, et~al., Journal of
  Physics G: Nuclear and Particle Physics \textbf{43}(8), 084001 (2016).
\newblock \doi{10.1088/0954-3899/43/8/084001}.
\newblock \urlprefix\url{http://dx.doi.org/10.1088/0954-3899/43/8/084001}

\bibitem{2019arXiv190805450B}
{Baikal-GVD Collaboration}, {:}, A.D. {Avrorin}, A.V. {Avrorin}, V.M.
  {Aynutdinov}, R.~{Bannash}, I.A. {Belolaptikov}, e.a. {Brudanin}, arXiv
  e-prints arXiv:1908.05450 (2019)

\bibitem{IC2018TXS}
M.~Aartsen, M.~Ackermann, J.~Adams, J.A. Aguilar, M.~Ahlers, M.~Ahrens,
  I.~Al~Samarai, D.~Altmann, K.~Andeen, et~al., Science \textbf{361}(6398),
  eaat1378 (2018).
\newblock \doi{10.1126/science.aat1378}.
\newblock \urlprefix\url{http://dx.doi.org/10.1126/science.aat1378}

\bibitem{Ciolfi2011}
R.~{Ciolfi}, S.K. {Lander}, G.M. {Manca}, L.~{Rezzolla}, Astrophys. J. Lett.
  \textbf{736}(1), L6 (2011).
\newblock \doi{10.1088/2041-8205/736/1/L6}

\bibitem{Guetta2020}
D.~{Guetta}, R.~{Rahin}, I.~{Bartos}, M.~{Della Valle}, \mnras \textbf{492}(1),
  843 (2020).
\newblock \doi{10.1093/mnras/stz3245}

\bibitem{AyalaSolares:2019iiy}
H.A. Ayala~Solares, et~al., Astropart. Phys. \textbf{114}, 68 (2020).
\newblock \doi{10.1016/j.astropartphys.2019.06.007}

\bibitem{CTA_book}
{Cherenkov Telescope Array Consortium, B.S.~Acharya et al.}, \emph{{Science
  with the Cherenkov Telescope Array}} (2019).
\newblock \doi{10.1142/10986}

\bibitem{2019Natur.575..464A}
{H. Abdalla et al. (H.E.S.S. Collaboration)}, Nature \textbf{575}(7783), 464
  (2019).
\newblock \doi{10.1038/s41586-019-1743-9}

\bibitem{2019Natur.575..455M}
{V.A. Acciari (MAGIC Collaboration)}, Nature \textbf{575}(7783), 455 (2019).
\newblock \doi{10.1038/s41586-019-1750-x}

\bibitem{2019Natur.575..459M}
{V.A. Acciari (MAGIC Collaboration)}, Nature \textbf{575}(7783), 459 (2019).
\newblock \doi{10.1038/s41586-019-1754-6}

\bibitem{2017ApJ...850L..22A}
{H. Abdalla et al. (H.E.S.S. Collaboration)}, Astrophysical Journal Letters
  \textbf{850}(2), L22 (2017).
\newblock \doi{10.3847/2041-8213/aa97d2}

\bibitem{2020ApJ...894L..16A}
{H. Abdalla et al. (H.E.S.S. Collaboration)}, Astrophysical Journal Letters
  \textbf{894}(2), L16 (2020).
\newblock \doi{10.3847/2041-8213/ab8b59}

\bibitem{2019MNRAS.489.2104T}
E.~{Troja}, A.J. {Castro-Tirado}, J.~{Becerra Gonz{\'a}lez}, Y.~{Hu}, G.S.
  {Ryan}, S.B. {Cenko}, R.~{Ricci}, G.~{Novara}, R.~{S{\'a}nchez-R{\'a}mirez},
  J.A. {Acosta-Pulido}, K.D. {Ackley}, M.D. {Caballero Garc{\'\i}a}, S.S.
  {Eikenberry}, S.~{Guziy}, S.~{Jeong}, A.Y. {Lien}, I.~{M{\'a}rquez}, S.B.
  {Pandey}, I.H. {Park}, T.~{Sakamoto}, J.C. {Tello}, I.V. {Sokolov}, V.V.
  {Sokolov}, A.~{Tiengo}, A.F. {Valeev}, B.B. {Zhang}, S.~{Veilleux}, MNRAS
  \textbf{489}(2), 2104 (2019).
\newblock \doi{10.1093/mnras/stz2255}

\bibitem{2020arXiv201207193M}
{V.A. Acciari (MAGIC Collaboration)}, arXiv e-prints arXiv:2012.07193 (2020)

\bibitem{allers13}
K.N. {Allers}, M.C. {Liu}, Astrophysical Journal \textbf{772}(2), 79 (2013).
\newblock \doi{10.1088/0004-637X/772/2/79}

\bibitem{Kirkpatrick99}
J.D. {Kirkpatrick}, I.N. {Reid}, J.~{Liebert}, R.M. {Cutri}, B.~{Nelson}, C.A.
  {Beichman}, C.C. {Dahn}, D.G. {Monet}, J.E. {Gizis}, M.F. {Skrutskie},
  Astrophysical Journal \textbf{519}(2), 802 (1999).
\newblock \doi{10.1086/307414}

\bibitem{Manjavacas20}
E.~{Manjavacas}, N.~{Lodieu}, V.J.S. {B{\'e}jar}, M.R. {Zapatero-Osorio},
  S.~{Boudreault}, M.~{Bonnefoy}, MNRAS \textbf{491}(4), 5925 (2020).
\newblock \doi{10.1093/mnras/stz3441}

\bibitem{Apai2013}
D.~{Apai}, J.~{Radigan}, E.~{Buenzli}, A.~{Burrows}, I.N. {Reid},
  R.~{Jayawardhana}, Astrophysical Journal \textbf{768}(2), 121 (2013).
\newblock \doi{10.1088/0004-637X/768/2/121}

\bibitem{Buenzli15}
E.~{Buenzli}, D.~{Saumon}, M.S. {Marley}, D.~{Apai}, J.~{Radigan}, L.R.
  {Bedin}, I.N. {Reid}, C.V. {Morley}, Astrophysical Journal \textbf{798}(2),
  127 (2015).
\newblock \doi{10.1088/0004-637X/798/2/127}

\bibitem{Yang2015}
H.~{Yang}, D.~{Apai}, M.S. {Marley}, D.~{Saumon}, C.V. {Morley}, E.~{Buenzli},
  {\'E}.~{Artigau}, J.~{Radigan}, S.~{Metchev}, A.J. {Burgasser}, S.~{Mohanty},
  P.J. {Lowrance}, A.P. {Showman}, T.~{Karalidi}, D.~{Flateau}, A.N. {Heinze},
  Astrophysical Journal Letters \textbf{798}(1), L13 (2015).
\newblock \doi{10.1088/2041-8205/798/1/L13}

\bibitem{Yang2016}
H.~{Yang}, D.~{Apai}, M.S. {Marley}, T.~{Karalidi}, D.~{Flateau}, A.P.
  {Showman}, S.~{Metchev}, E.~{Buenzli}, J.~{Radigan}, {\'E}.~{Artigau}, P.J.
  {Lowrance}, A.J. {Burgasser}, Astrophysical Journal \textbf{826}(1), 8
  (2016).
\newblock \doi{10.3847/0004-637X/826/1/8}

\bibitem{NRT}
C.M. Copperwheat, I.A. Steele, R.M. Barnsley, S.D. Bates, M.F. Bode, N.R. Clay,
  C.A. Collins, H.E. Jermak, J.H. Knapen, J.M. Marchant, C.J. Mottram, A.S.
  Piascik, R.J. Smith, in \emph{Ground-based and Airborne Telescopes VI}, vol.
  9906, ed. by H.J. Hall, R.~Gilmozzi, H.K. Marshall. International Society for
  Optics and Photonics (SPIE, 2016), vol. 9906, pp. 1176 -- 1183.
\newblock \doi{10.1117/12.2231755}.
\newblock \urlprefix\url{https://doi.org/10.1117/12.2231755}

\end{thebibliography}

%
%

\end{document}